\documentclass[10pt,journal,compsoc]{IEEEtran}
\ifCLASSOPTIONcompsoc
  \usepackage[nocompress]{cite}
\else
  \usepackage{cite}
\fi
\usepackage{cite}                      
\usepackage{amsmath,amsthm,amssymb}
\usepackage{xspace}
\usepackage{hyperref}
\usepackage{xcolor}
\usepackage{graphicx}                
\usepackage{enumitem}
\DeclareGraphicsExtensions{.pdf,.png,.jpg,.jpeg}
\usepackage{booktabs}                  
\newcommand{\myparagraph}[1]{\vspace{1mm} \noindent \textbf{#1}}
\newcommand{\ie}{i.e.,\xspace}
\newcommand{\etal}{et al.\xspace}

\ifCLASSINFOpdf
\else
\fi
\begin{document}
%
\title{Continuous Scatterplot Operators for Bivariate Analysis and Study of Electronic Transitions}
%
%
%
%

\author{Mohit Sharma,~\IEEEmembership{}
        Talha Bin Masood,~\IEEEmembership{}
        Signe S. Thygesen,~\IEEEmembership{}
        Mathieu Linares,~\IEEEmembership{}
        Ingrid Hotz,~\IEEEmembership{Member,~IEEE}
        and Vijay Natarajan,~\IEEEmembership{Member,~IEEE}
\IEEEcompsocitemizethanks{\IEEEcompsocthanksitem M. Sharma and V. Natarajan are with the Department of Computer Science and Automation, Indian Institute of Science, Bangalore.\protect\\
E-mail: \{mohitsharma, vijayn\}@iisc.ac.in.
\IEEEcompsocthanksitem T.B. Masood, S. S. Thygesen and M. Linares are with the Department of Science and Technology (ITN), Link\"oping University, Norrk\"oping, Sweden.\protect\\
E-mail: \{talha.bin.masood, signe.sidwall.thygesen, mathieu.linares\}@liu.se.
\IEEEcompsocthanksitem  I. Hotz is with the Department of Science and Technology (ITN), Link\"oping University, Norrk\"oping, Sweden, and Indian Institute of Science, Bangalore.\protect\\
E-mail: ingrid.hotz@liu.se.
}
\thanks{}}

%
%

\markboth{IEEE Transactions on Visualization and Computer Graphics,~Vol.~X, No.~X, Month~Year}%
{Sharma \MakeLowercase{\textit{et al.}}: Continuous Scatterplot Operators for Bivariate Analysis and Study of Electronic Transitions}
%



\IEEEtitleabstractindextext{%
\begin{abstract}
Electronic transitions in molecules due to the absorption or emission of light is a complex quantum mechanical process. Their study plays an important role in the design of novel materials. A common yet challenging task in the study is to determine the nature of electronic transitions, namely which subgroups of the molecule are involved in the transition by donating or accepting electrons, followed by an investigation of the variation in the donor-acceptor behavior for different transitions or conformations of the molecules. In this paper, we present a novel approach for the analysis of a bivariate field and show its applicability to the study of electronic transitions. This approach is based on two novel operators, the continuous scatterplot (CSP) lens operator and the CSP peel operator, that enable effective visual analysis of bivariate fields. Both operators can be applied independently or together to facilitate analysis. 
The operators motivate the design of control polygon inputs to extract fiber surfaces of interest in the spatial domain. The CSPs are annotated with a quantitative measure to further support the visual analysis. We study different molecular systems and demonstrate how the CSP peel and CSP lens operators help identify and study donor and acceptor characteristics in molecular systems.
\end{abstract}

\begin{IEEEkeywords}
Bivariate field analysis, Continuous scatterplot, Fiber surface, Control polygon, Visual analysis, Electronic transitions.

\end{IEEEkeywords}}

\maketitle

\IEEEdisplaynontitleabstractindextext

%
\IEEEpeerreviewmaketitle

\IEEEraisesectionheading{\section{Introduction}\label{sec:introduction}}

%
%
%
%
\IEEEPARstart{T}{he} study of electronic transitions in molecules due to absorption or emission of light is crucial for understanding their chemical and physical properties~\cite{Kim2019}. Electrons in a molecule are distributed within a series of available orbitals~\cite{Mulliken}. These electrons are excited from occupied orbitals to unoccupied orbitals when the molecule absorbs a photon, resulting in a change in the electronic structure of the molecule. The transition is reversed upon emission of light.  The Natural Transition Orbital (NTO) is a compact representation of electronic excitations. It considers a linear combination of orbitals involved in a specific electronic transition to describe from where the electrons are excited  (hole NTO) and to where they are promoted (particle NTO)~\cite{Martin2003NTO}. 

\begin{figure*}[!t]
{
  \centering
  \includegraphics[width=\linewidth]{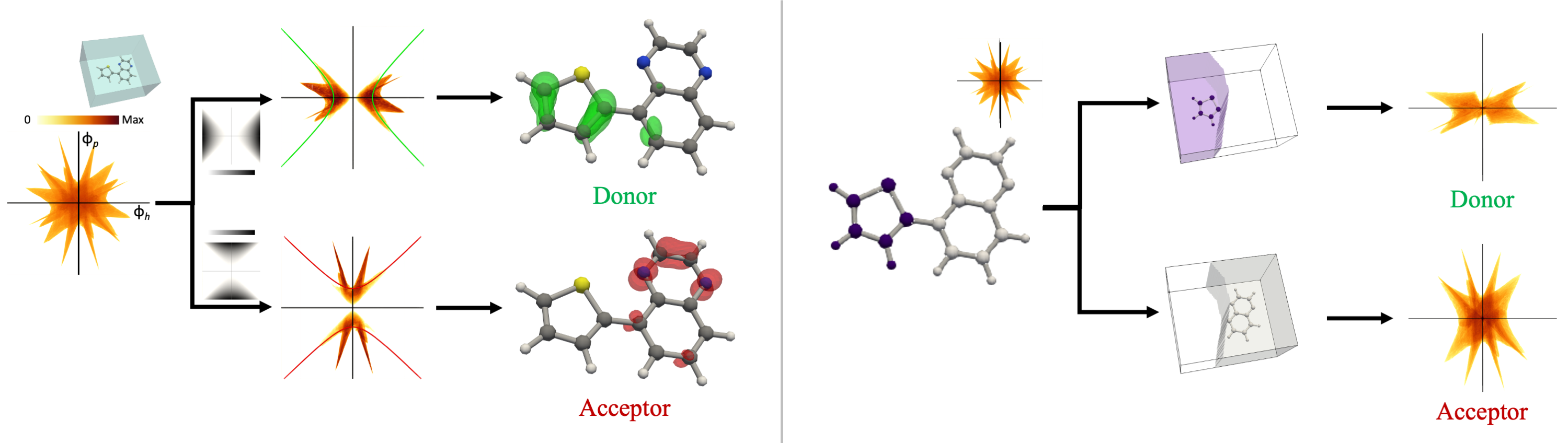}
  \caption{Studying electronic transitions in a Thiophene Quinoxaline molecule. (Left)~CSP lens operator allows to query the range space, \ie CSP of the hole and particle NTO ($\phi_h$, $\phi_p$), directly to extract the donor (top) and acceptor (bottom) regions. Together with a fiber surface extraction step, this operator provides a way to link views of range space and the spatial domain. Fiber surfaces corresponding to the query control polygons (green and red) are displayed on a ball-and-stick model of the molecule. (Right)~Domain segmentation driven CSP peel operator allows to move from the spatial domain to range space. Peeled CSP corresponding to the Thiophene and Quinoxaline subgroups, obtained using a Voronoi segmentation, help study their behavior.}
  \vspace{-1.2em}
  \label{fig:teaser}
}\end{figure*}

Our objective in this paper is to develop a solution for a problem of interest to a theoretical chemist, namely the analysis of the nature of electronic transitions in molecules. A typical study of electronic transitions poses questions related to the change in charge localization upon excitation between the different molecular subgroups such as:
Which molecular subgroup is a donor / acceptor for a particular excited state of the molecule? 
How do the donor / acceptor strengths (measured based on the amount of charge transferred) vary with differing molecular conformations? 
How do donor / acceptor behavior vary within a family of complexes?
Can a given state be classified as presenting a Local Excitation (within a subgroup) or a Charge Transfer Excitation (between subgroups) character? 
To answer these questions, we present a new approach to analyze a bivariate field and demonstrate its effectiveness by applying it towards the study of electronic transitions in different molecules. In contrast to existing methods that study the charge density fields individually, we propose the study of a bivariate field consisting of two scalar fields, the hole NTO and the particle NTO. 
The continuous scatterplot (CSP) is a generalization of the discrete scatter plot to continuous multivariate fields defined on $n$-dimensional spatial domains~\cite{Bachthaler2008CSP}. We introduce two independent operators, the CSP lens operator and the CSP peel operator, for analyzing a bivariate field. \autoref{fig:teaser} illustrates the visual analysis of a simple molecule using the two operators. The CSP lens operator directly queries the range space to identify characteristics of interest, for example to extract the donor and acceptor regions. Subsequently, fiber surfaces\cite{carr2015fiber}  are computed to visualize the corresponding spatial regions. The CSP peel operator is directed by a segmentation of the spatial domain of the bivariate field. It peels CSP layers based on individual segments of the domain that correspond to molecular subgroups. Subsequently, the peeled CSP layers may be compared or studied individually.


\subsection{Related  Work}
Electron density distribution in a molecule is often visualized via isosurfaces together with a ball-and-stick model representation of the molecule and additional annotations to indicate transfer of charge~\cite{Humphrey1996,Stone2011}.  These techniques are useful for the study of  orbitals within a given state. The study of transitions based on these techniques rely on side-by-side comparisons~\cite{Haranczyk2008}. The analysis of transfer of charge between molecular subgroups and their classification as donor or acceptor is a key step towards the study of electronic transitions. A few studies employ quantitative approaches towards the study of charge transfer. Garcia~\etal~\cite{Garcia2010} and Bahers~\etal~\cite{Bahers2011} propose a set of indexes based on pointwise difference between the charge density fields to capture the amount of charge transfer and change in dipole moment. An extension of these indices may be interpreted as a hole-electron distance~\cite{Guido2013,Huet2020}. 

In recent work, Masood~\etal~\cite{masood2021visual} present an automated method for quantifying charge distribution and transitions based on a spatial segmentation of the electronic density field. They formulate charge transfer as a constrained optimization problem by modeling the molecular subgroups as nodes in a graph and describe methods for visualizing the charge distribution and transfer at the atomic and subgroup level.
An approach to extend this to an ensemble of electronic transitions is presented by Thygesen~\etal~\cite{Thygesen2022}. They propose a clustering-based pipeline for interactive visual analysis and exploration.
Previous work, as described above, has either focused on visual comparisons of the individual charge density fields or on methods to compute the electronic charge transferred between molecular subgroups and to visualize these quantities. In contrast, we present an approach based on the analysis of a bivariate field to study the transitions. 

Generalization of discrete scatter plots~\cite{sarikaya2017scatterplots} to continuous scalar fields~\cite{Bachthaler2008CSP} and of isosurfaces to fiber surfaces~\cite{carr2015fiber} has led to new directions in bivariate and multivariate field visualization.  The CSP generalizes discrete scatter plots to spatially continuous data, namely multivariate fields defined on an $n$-dimensional spatial domain~\cite{Bachthaler2008CSP}. It presents a dense visualization and scales well for large data sizes. Follow-up work on discontinuities in the CSP establishes a relationship with the number of connected components in the spatial domain and demonstrates its applicability towards the visual analysis of bivariate fields~\cite{Lehmann2010}. The study of relationship between the individual scalar fields has been used to identify interesting isovalues~\cite{nagaraj2010relation} in the multivariate setting, following the use of isosurface statistics in the case of univariate fields~\cite{hamish2006histograms}. 
Carr~\etal~\cite{carr2015fiber} define fibers as a generalization of isosurfaces to bivariate fields and introduce the fiber surface, a separating surface which consists of a collection of fibers. They  demonstrate the applicability of fiber surface towards the study of  electron density and its derived fields, and for characterization of bonds in molecules. Subsequent work focused on improving the correctness and efficiency of fiber surface computation~\cite{klacansky2016fast} and various applications~\cite{tierny2016jacobi,Blecha2019Nuclear,Raith2019Tensor}. Tierny \etal~\cite{tierny2016jacobi}  segment the domain based on the Reeb space~\cite{edelsbrunner2008reeb} and use them to peel the CSP to reveal its connected structures. 

This paper presents a generic approach for visual analysis of bivariate fields that utilizes the CSP and fiber surface representations. Spatial distribution of the amount of charge donated and gained are the two main quantities needed to study an electronic transition. The orientation, density, and other shape characteristics in the plotted CSP provide a visual representation of the relationship between these two quantities. Subsequently, fiber surfaces are used to extract the corresponding spatial regions. 


\subsection{Contributions}
We introduce a novel approach for visual analysis of bivariate fields  motivated by the objective of developing effective methods to support the study of electronic transitions. We employ techniques including CSP and fiber surfaces to develop methods for exploring the bivariate field both within the spatial domain and in the range space.  Key contributions of this work include
\begin{enumerate}[noitemsep,topsep=0pt,leftmargin=0.4cm]
    \item A CSP lens operator that enables queries on the range space based on application-specific characteristics.  
    \item A CSP peel operator that enables the extraction of different layers of the CSP based on a user-specified domain segmentation.
    \item A control polygon selection process that is directed by the CSP lens operator and helps in the extraction of fiber surfaces of interest. 
    \item A visual analysis pipeline that utilizes the two operators independently or conjointly. 
    \item A quantification of CSPs to support visual analysis.
    \item Case studies on two molecular systems that demonstrate the utility and effectiveness of the pipeline in supporting the study of electronic transitions.
\end{enumerate}

\begin{figure*}[!t]
    \centering
    \includegraphics[width=\textwidth]{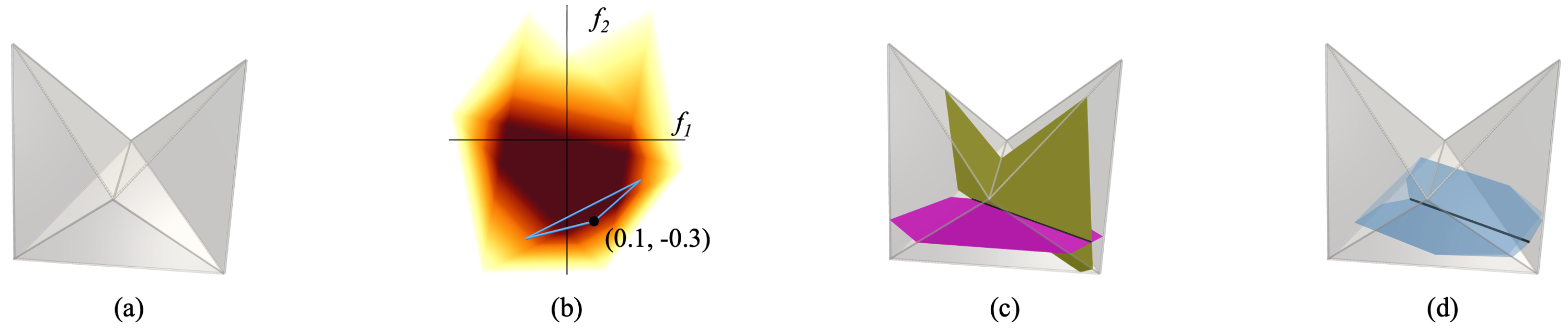}
    \caption{(a)~A synthetic bivariate field, $f_1$ is the x-coordinate and $f_2$ is the y-coordinate. (b)~CSP of the bivariate field. A blue control polygon is selected, it contains the black point $(0.1, -0.3)$. (c)~The olive color isosurface of $f_1$ corresponds to isovalue 0.1 and the pink isosurface of $f_2$ corresponds to isovalue -0.3. The two isosurfaces intersect along the black line segment, the fiber corresponding to $(0.1, -0.3)$. (d)~The blue fiber surface corresponding to the control polygon contains the black fiber.} 
    \vspace{-1.2em}
    \label{fig:background}
\end{figure*}

A previous paper presented the CSP peel operator and its application to the study of electronic transitions in two molecular systems~\cite{sharma2021segmentation}. In this extended version, we additionally present the CSP lens operator that incorporates a new dimension to the visual analysis pipeline, a detailed description of how the two operators can be applied independently and in unison, new quantification methods to support the understanding of the output of the operators, the use of fiber surfaces to visualize the output of the CSP lens operator, and additional case studies that demonstrate the utility of the extended pipeline.

\section{Background}\label{sec:background}
This section provides a brief introduction to techniques for bivariate field analysis and visualization, and a primer on electron density fields, orbitals, and two molecular systems that are the focus of our study. 

\myparagraph{Bivariate field.} A multivariate field is a set of scalar valued functions defined over a spatial domain. A \emph{bivariate field} is a specific instance when two scalar fields are defined over the domain $\mathcal{M}$,
\begin{align*}
f = \{f_1,f_2\} : \mathcal{M} \rightarrow \mathbb{R}^2.
\end{align*}
The two scalar fields $f_1$ and $f_2$ may either be independent of each other or be related. Studying the bivariate field helps identify the relationship between the two fields in addition to understand the characteristics of individual scalar fields. 

\autoref{fig:background}(a) shows a synthetic data set from the topology toolkit \textsc{ttk}~\cite{Tierny2018ttk}. The simply connected volumetric domain is represented by a set of tetrahedra, $f_1$ is the x-coordinate and $f_2$ is the y-coordinate of the point in $\mathcal{M}$.

\myparagraph{Continuous scatterplot.} A CSP is a density plot in the range space, generalizing the concept of a scatterplot to spatially continuous multivariate data~\cite{Bachthaler2008CSP}. It is closely related to histograms and isosurface statistics~\cite{hamish2006histograms, scheidegger2008revisiting}. The density at a point $(f_1 = s_1,f_2 = s_2)$ in the range space, called the point density, is mapped to color. \autoref{fig:background}(b) shows the CSP of the synthetic bivariate field. The CSP is a good representation to identify regions of interest within the spatial domain and serves as a useful interface to select these regions.

\myparagraph{Fiber.} A fiber~\cite{carr2015fiber} is a generalization of the isosurface to bivariate fields. Given a pair of scalar values $(s_1,s_2)$, a \emph{fiber} $\mathcal{F}$  represents the collection of points in $\mathcal{M}$ that are mapped to $(s_1,s_2)$ under $f$, $\mathcal{F} = f^{-1}(s_1, s_2)$. \autoref{fig:background}(c) shows the isosurface of $f_1^{-1}(s_1)$ in olive color and that of $f_2^{-1}(s_2)$ in pink. The fiber is the intersection of the two isosurfaces, shown in black. The black point in the CSP shown in \autoref{fig:background}(b) corresponds to the black fiber in \autoref{fig:background}(c).

\myparagraph{Fiber surface.} The preimage of a collection of points from the CSP that lie on a continuous curve in the range space is a collection of fibers. This collection of fibers forms a \emph{fiber surface}.  The blue translucent surface in \autoref{fig:background}(d) is obtained as the preimage of the edges of the blue triangle in  \autoref{fig:background}(b). Note that the black fiber lies within the fiber surface.

\myparagraph{Control polygon.} The blue triangle is called a \emph{control polygon} because it serves as a tool to select interesting regions of the CSP and visualize the corresponding fiber surface in the spatial domain. The control polygon may be open or closed polygon, or even a line segment as required by the query. The connectivity of the corresponding fiber surface changes accordingly.

\myparagraph{Electronic transition and orbital.} A molecular orbital is a wave function that describes the location and wave behavior of an electron in a molecule. An electron density field, which represents the probability of an electron being present at a given location,  can be derived from the orbital. Upon absorption of light, the electronic structure of the molecule is affected as the electrons are excited from occupied to unoccupied orbitals.
Chemists use a compact representation for an electronic transition known as Natural Transition Orbital~(NTO)~\cite{Martin2003NTO}, which is defined as a linear combination of orbitals that are involved in a specific electronic transition. There are two types of NTOs: the hole NTO ($\phi_h$) which represents the orbital vacated by an electron during an excitation, and the particle NTO ($\phi_p$) which corresponds to the orbital occupied by the electron. 
We analyze the bivariate field ($\phi_h,\phi_p$) consisting of both NTOs in order to study electronic transitions in molecules. 

\begin{figure*}[!t]
    \centering
   \includegraphics[width=\textwidth]{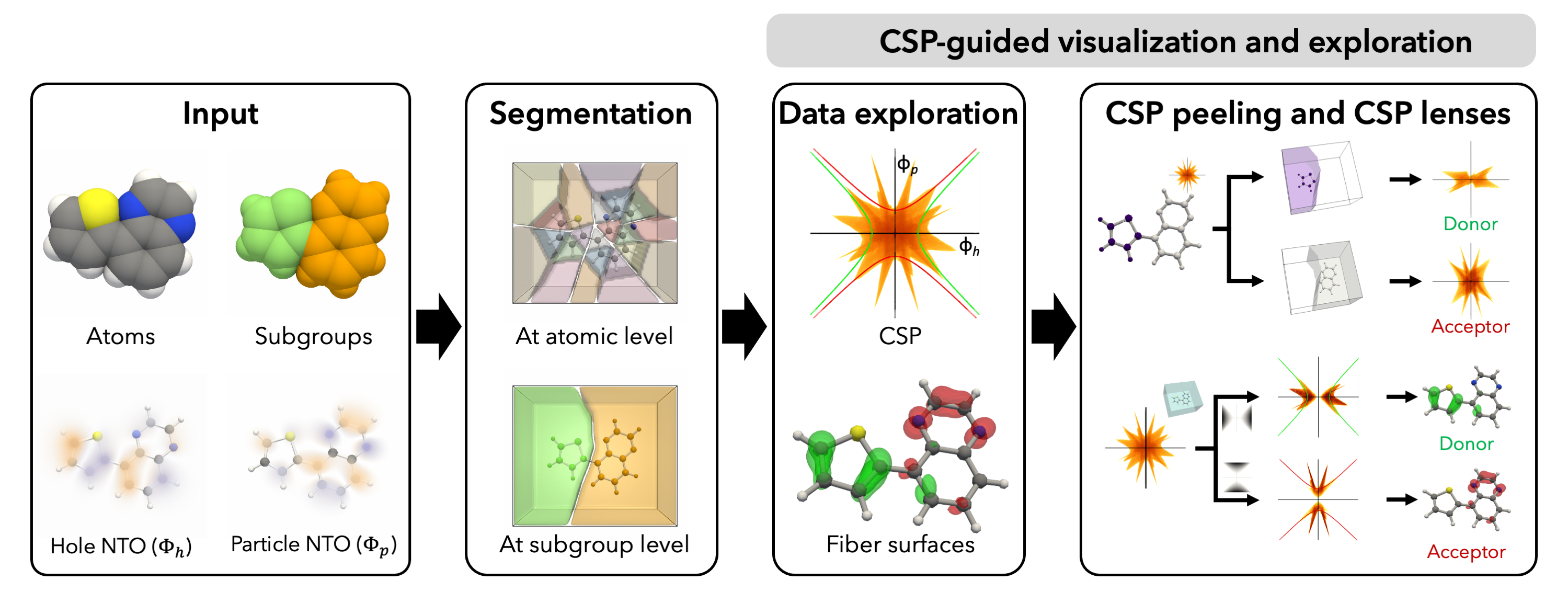}
    \caption{Visual analysis workflow. Molecular structure, NTO fields, and subgroup descriptions are available as input. A weighted Voronoi segmentation identifies atomic regions. The segment corresponding to a molecular subgroup is computed as the union of atomic regions.
The CSP of the bivariate field may be explored either using fiber surfaces or CSP peel and lens operator can be used for a systematic analysis and donor-acceptor strength comparisons. The peeled CSPs can be visually analyzed for donor and acceptor behavior. The lens operator globally extracts donor and acceptor regions throughout the molecule. For all the CSPs presented in this paper, we have used consistent color map (\includegraphics[width=0.07\textwidth]{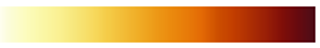}) with log scale mapping.
} 

\vspace{-1.2em}
    \label{fig:pipeline}
\end{figure*}

\myparagraph{Molecular system.} In this paper, we consider two case studies: Thiophene-Quinoxaline (TQ) and copper complexes. The molecule is considered as a collection of multiple subgroups. A \emph{subgroup} in a molecular system is a collection of atoms that behave as a single entity in the electronic transitions study. The subgroup information is provided by the chemist.
TQ is composed of two subgroups, namely thiophene and quinoxaline, that are used in the field of organic electronics either independently or together to create conductive polymers. Thiophene is well known for donating electrons whereas Quinoxaline accepts electrons. \autoref{fig:teaser} shows the TQ system. The five-member ring on the left is the Thiophene subgroup, Quinoxaline is on the right. We consider different geometric conformations of this molecule in our study. The second molecular system that we study is a set of copper complexes. The complexes present variations in the ligand. All copper complexes contain three subgroups as shown in the top row of \autoref{fig:cu-ligands}: Copper (Cu) in the center, Phenanthroline (PHE) as a fixed subgroup, and a third subgroup that varies between Phenanthroline (PHE), Dimethoxy Phenanthroline (PHEOME), and a Diphosphine ligand (XANT). 

\section{Bivariate  analysis  of  electronic  transitions}\label{sec:bivariate}
The CSP of the bivariate field ($\phi_{h}, \phi_{p}$) may be queried using fiber surfaces to identify regions that have a strong donor or acceptor characteristic.
Identification of control polygons that correspond to strong donor or acceptor behavior is a continual challenge. A detailed visual inspection of the CSP is  necessary to select an interesting control polygon and the selection often succeeds only after multiple attempts. This difficulty motivated the development of a technique and a tool to query the CSP directly and extract meaningful regions. We call this range-based query operator as a CSP lens. A CSP lens reduces the size of the search space and helps identify a region of interest. An example for the TQ molecular system, with its unique star-shaped CSP, is shown in \autoref{fig:teaser} (left). We queried the donor and acceptor regions of CSP using donor and acceptor lenses. Selecting a control polygon in the donor CSP results in a fiber surface that represents the donor points in the spatial domain.  Two pairs of piecewise linear curves (red and green) across the spikes of the CSP are selected as control polygons.

The green fiber surface extends across the Thiophene subgroup (donor) and the red fiber surface extends across the Quinoxaline subgroup (acceptor). These observations and further exploration of the CSP resulted in the insight that the behavior of a subgroup as donor or acceptor is encoded as patterns within the CSP, which may serve as a descriptor for electronic transitions. This motivated the idea of a CSP peel operator. We propose a smart peel operator that is applied on a segmentation of the spatial domain to generate CSPs corresponding to individual subgroups. This process can be imagined as peeling away different layers from the CSP to reveal contributions of individual molecular subgroups of interest. The CSP peel operator provides direct access to transition patterns of individual atoms or atomic groups thereby simplifying the exploration. It displays the contributions of a select set of subgroups to the CSP and enables their comparison. The example in \autoref{fig:teaser} is simple, with dominant and distinct donor and acceptor regions along the two axes. The output of the peel operator may be complex and require a deeper study but they do support systematic exploration. We consistently use the same axes and color map (log scale) for the CSPs in all figures within this paper. 

\section{Visual analysis pipeline}
\autoref{fig:pipeline} presents an overview of the visual analysis pipeline that processes the input bivariate field together with application-specific information that determine a spatial domain segmentation. For the application to the study of electronic transitions, the bivariate field consists of the pair of NTOs and the molecular structure information (atom locations, radii, bond information) is available as input.

\begin{figure*}[!t]
    \centering
    \includegraphics[width=\textwidth]{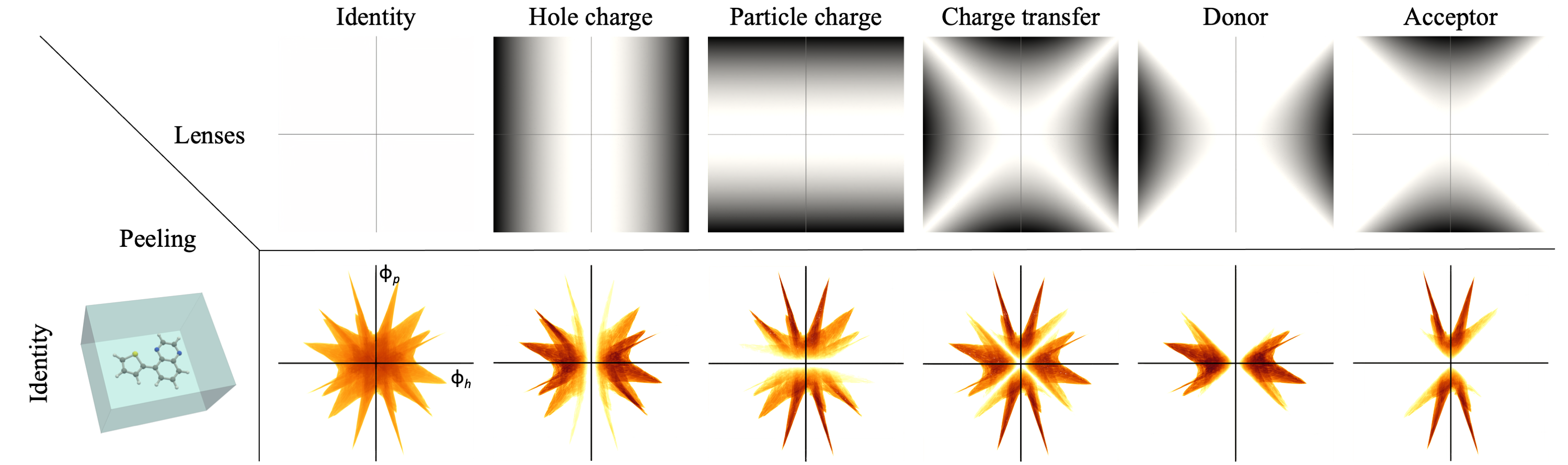}
    \caption{CSP lens operators applied on the TQ molecular system. Top row shows the mask associated with the different lens operators. Bottom row shows the results after applying a lens operator to the CSP of the bivariate field ($\phi_{h}, \phi_{p}$) over the entire molecule. All lenses are designed to exclude the regions close to origin.} 
    \vspace{-1.2em}
    \label{fig:csp_lenses}
\end{figure*}

\subsection{CSP lens operator}
\label{sec:lenses}
A common challenge in the analysis of a bivariate field is the identification of interesting patterns in the CSP that can help extract the desired spatial regions via fiber surfaces. We introduce CSP lens operator as the first method in the CSP-guided visualization and exploration stage of our proposed visual analysis pipeline. The CSP lens can directly query a CSP in an automated manner as compared to visually identifying patterns. A CSP lens operator is specified using a mask over the CSP with varying weight. The weight of each point in the mask depends on the specific query.  \autoref{fig:csp_lenses} shows the mask associated with different lenses on the x-axis with a focus on the study of electronic transitions. Donor, acceptor, and other lens queries translate to different mathematical expressions that define the mask, as shown in the following list.  In each expression, $N$ denotes the total number of points in a CSP. 
\begin{align*}
Identity\ lens = 1_i \quad \forall i \in [0,N-1]\\
Hole\ charge\ lens\ (H_L) = |\phi_{h_i}|^2 \quad \forall i \in [0,N-1]\\
Particle\ charge\ lens\ (P_L) = |\phi_{p_i}|^2 \quad \forall i \in [0,N-1]\\
Charge\ transfer\ lens = |H_L - P_L| \\
Donor\ lens = \max(H_L - P_L,0) \\
Acceptor\ lens = \max(P_L - H_L,0) 
\end{align*}
Let $d_i$ denote the point density at each point $i$ of the CSP. The CSP lens operator is defined as the pointwise product of the mask and point density. For example, the hole charge lens operator results in a hole charge CSP: 
\begin{align*}
Hole\ charge\ CSP = d_i \cdot |\phi_{h_i}|^2 \quad \forall i \in [0,N-1]
\end{align*}

\myparagraph{Interpretation.} 
The result of the application of a CSP lens operator is also a CSP.  The CSP lens operators shown in \autoref{fig:csp_lenses} for the application to electronic transitions partition the CSP into donor and acceptor regions. For example, the charge transfer lens computes the amount of charge transferred to or from a particular point by calculating the difference between particle NTO and hole NTO. The donor lens extracts the region within the CSP where charge lost is positive and the acceptor lens extracts the regions where charge gained is positive. Also note that, all the listed lens operators exclude the region near origin by using low values in the mask. The points close to origin map to very low hole NTO and particle NTO values. They correspond to spatial points that lie far from atoms and near the boundary of the domain. Selecting a control polygon near the origin produces a fiber surface that covers the complete molecule. We are interested in analyzing the behavior of subgroups and regions close to atoms. As we move away from the vicinity of atoms, both the amount of charge donated and charge gained decrease. Points in the CSP that are close to origin map to such regions in the domain and hence may not be useful for analysis. Excluding this region in the CSP reduces the search space.

The CSP lens operator is applicable generically for all bivariate data analysis and not limited to the application to electronic transitions. The operator needs to be specified in a way that it segments the range space corresponding to desired features. In the electronic transitions application, the chemist is interested in locating the donor and acceptor regions to understand a particular transition. In this specific case, the mask may be defined using precise mathematical expressions. The visual analysis pipeline only requires that the mask be expressed as a scalar function over the range space.

\begin{figure}[!t]
    \centering
   \includegraphics[width=0.5\textwidth]{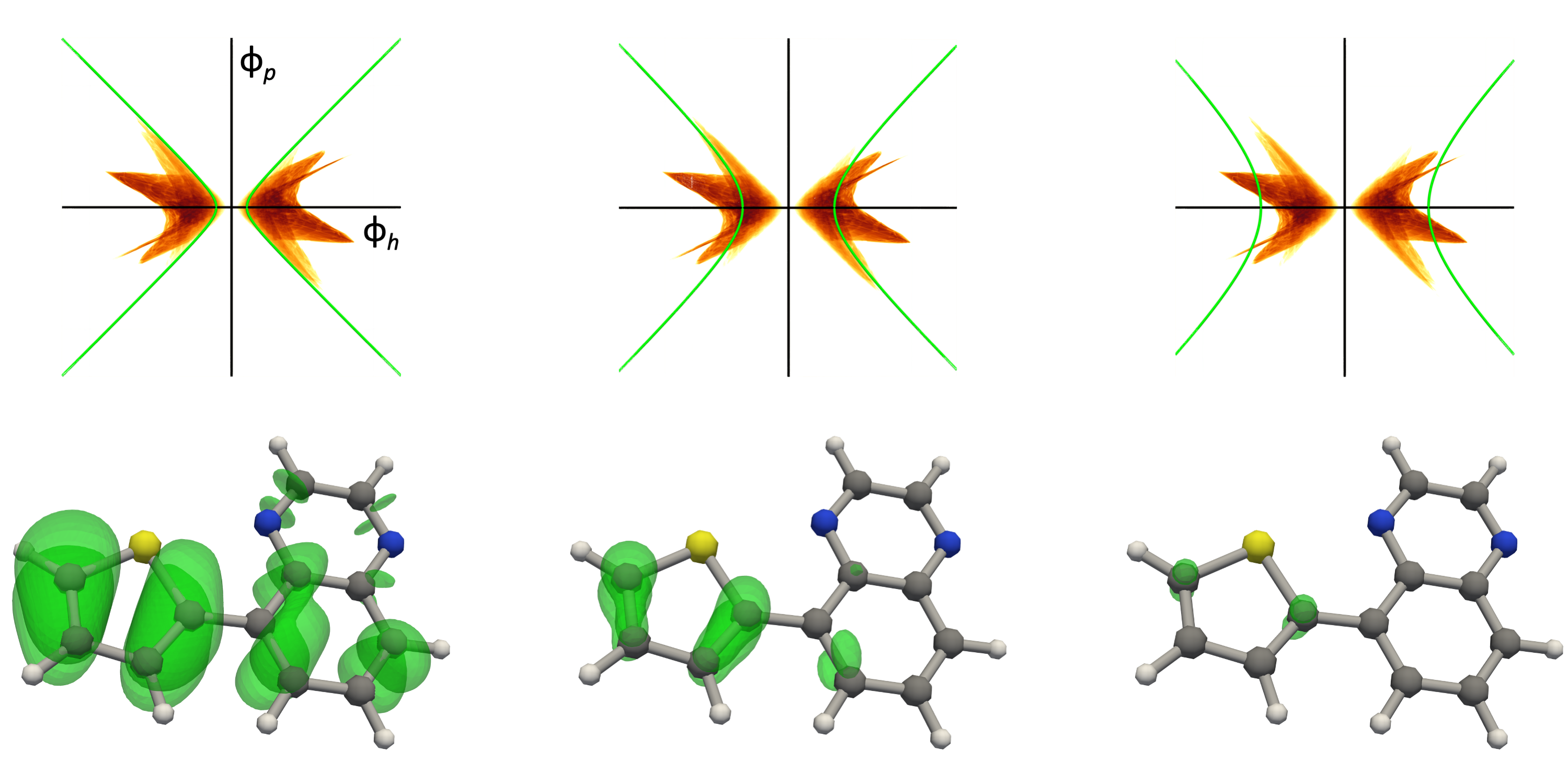}
    \caption{Exploring the CSP using fiber surfaces. As the control polygons (green curves, top row; left to right) move away from origin, the corresponding fiber surfaces (green, bottom row; left to right) shrink and  move closer to the atoms. The control polygons in the rightmost CSP represent stronger donors as compared to the others. The corresponding fiber surfaces highlight the spatial points that donate larger amounts of charge as compared to others.} 
    \vspace{-1.2em}
    \label{fig:cp_variation}
\end{figure}

\subsection{CSP lens and fiber surface}

The control polygon selection process may be directed by the output of the lens operator. The utility of the CSP lens operator increases with a visualization of fiber surfaces that correspond to the selected control polygons.  
 
\myparagraph{Control polygon selection.}  The shape, location, and orientation of control polygons depends largely on the objective of the study. The CSP lens operator helps direct attention towards certain regions and hence reduce the search space for control polygon selection. For example,  \autoref{fig:cp_variation} shows the output of the donor lens for a particular molecular conformation. A control polygon traced within the nonempty regions of  this CSP would extract a donor fiber surface. The reduction in search space helps in smart placement of control polygons. In the examples or case studies that we consider, we wish to compare spatial locations on basis of their donor and/or acceptor strength. We select control polygons such that they correspond to fiber surfaces representing a specific donor or acceptor strength $k$ \ie points on the spatial domain satisfying $|\phi_{h}|^2 - |\phi_{p}|^2  = k$. Within the 2D range space, this equation defines the curve (isocontour) shown in \autoref{fig:cp_variation}, a hyperbola with origin as center and $k$ as semi major axis. The corresponding fiber surface may be considered as the isosurface of a donor strength ($|\phi_{h}|^2 - |\phi_{p}|^2$) field and the control polygon is an isocontour of the mask function associated with the donor lens operator for isovalue $k$. The CSP is sampled at vertices of a 2D grid. Hence, the isocontours are represented as piecewise linear curves and serve as a control polygon (often open) for fiber surface computation~\cite{klacansky2016fast}. In the rest of the paper, we use the term control polygon to refer to these isocontours of the mask function. A control polygon that lies close to origin corresponds to weaker donor fiber surfaces when compared to control polygons that lie further away from origin. 
We observe an analogous behavior for the acceptor lens. In general, a lens represents a specific characteristic. We may associate one or more template control polygons with a lens operator, for example based on isocontour as discussed above or based on an application-specific characteristic. 
 
\begin{figure}[!t]
    \centering
    \includegraphics[width=0.5\textwidth]{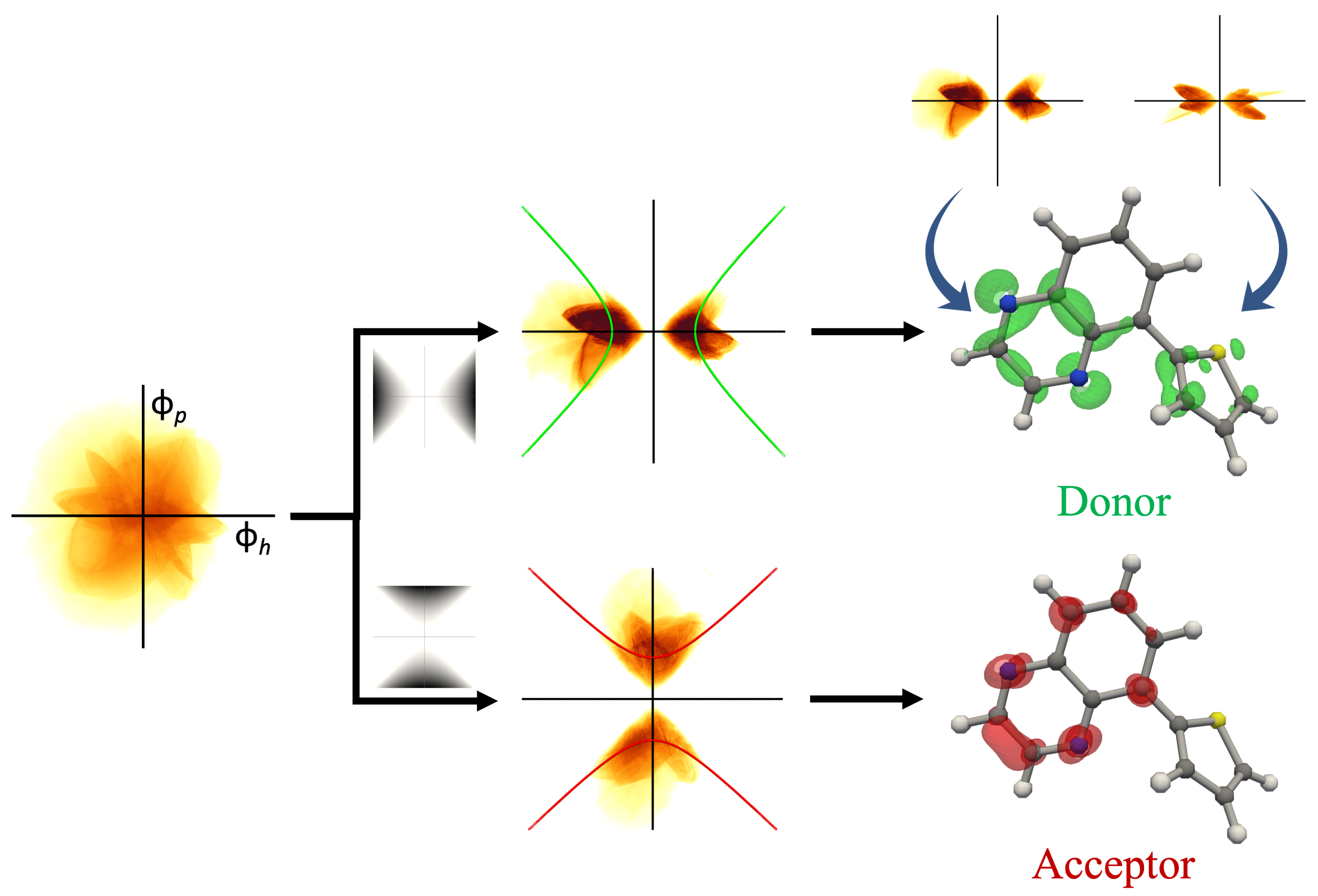}
    \caption{Donor and acceptor regions extracted using CSP lens operator and corresponding fiber surfaces for TQ-$120^{\circ}$. Quinoxaline behaves as both donor and acceptor in this conformation. Insets show the donor CSP layers corresponding to both subgroups.}
    \vspace{-1.2em}
    \label{fig:tq-120-lens}
\end{figure}

\begin{figure*}[!t]
    \centering
   \includegraphics[width=\textwidth]{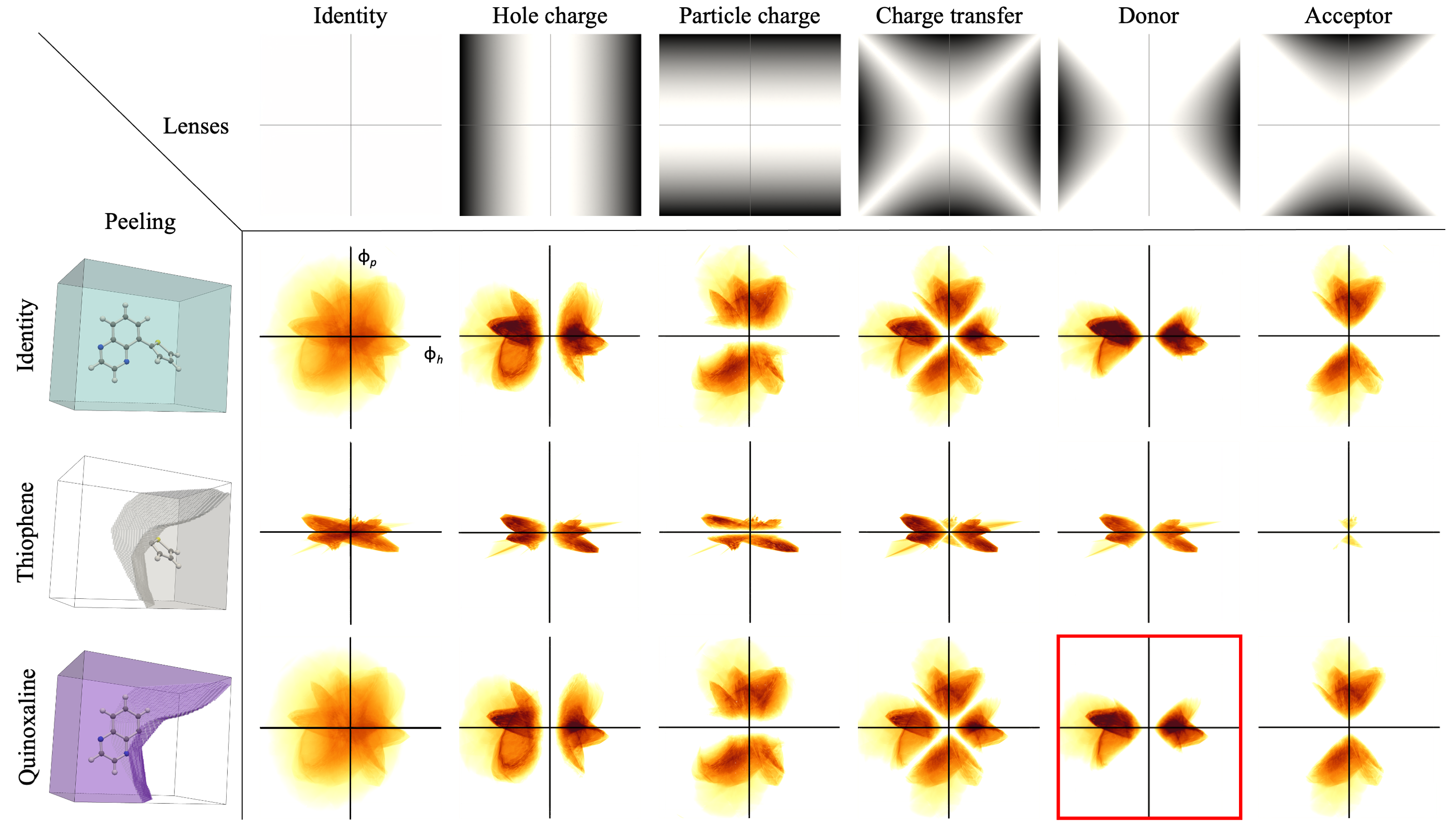}
    \caption{CSP lens operator based exploration along X-axis and CSP peel operator based exploration along Y-axis. Joint application of the two operators enables extraction of donor and acceptor CSPs corresponding to individual subgroups. The highlighted (red) CSP is the donor CSP specific to the Quinoxaline subgroup.} 
    \vspace{-1.2em}
    \label{fig:csp_lens-peel}
\end{figure*}

 \myparagraph{Motivation for a second operator.} The CSP lens operator executes the given query globally, over the entire CSP. In the absence of additional information, the lens operator supports queries only along one axis (x-axis in \autoref{fig:csp_lens-peel}), namely on the range space. We incorporate a second axis of exploration to support further analysis of the bivariate field. This second axis of exploration is based on queries on the spatial domain, specifically a segmentation of the domain. The donor (green) region and acceptor (red) region in \autoref{fig:teaser} (left) are well separated. The donor belongs to the single five-member ring Thiophene and acceptor belongs to Quinoxaline. The molecule in consideration is a simple two subgroup system in which one subgroup acts as donor while the other acts as acceptor for this specific geometric conformation. The molecular system in \autoref{fig:tq-120-lens} is a different conformation of the same molecule where Thiophene acts as pure donor while Quinoxaline acts as both donor and acceptor. The fiber surfaces generated from donor CSP lens operator output cover both Thiophene and Quinoxaline subgroups. It implies that the donor CSP consists of two overlapping layers, one from each subgroup (Insets in \autoref{fig:tq-120-lens}). If we wish to individually highlight the two donor fiber surfaces of Thiophene and Quinoxaline, then we need an additional operator that partitions the CSP  based on a segmentation of the domain or, more specifically, peel the overlapping CSP layers. 

\begin{figure*}[!t]
    \centering
   \includegraphics[width=\textwidth]{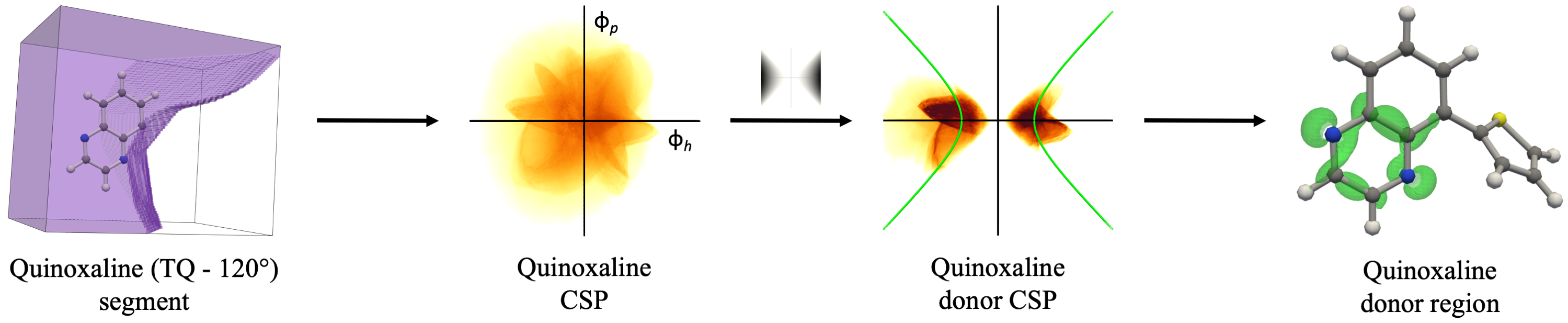}
    \caption{CSP peel and lens operator applied simultaneously to extract subgroup specific donor region. The domain segmentation for Quinoxaline generates its peeled CSP layer. The donor lens operator may be applied within this peeled layer. Further, selecting a control polygon in the donor CSP helps extract the donor fiber surface restricted to Quinoxaline.} 
    \vspace{-1.2em}
    \label{fig:lens-illustration}
\end{figure*}

\subsection{CSP peel operator}\label{sec:csp_peel}
We now describe a second operator for analyzing the CSP via extraction of different layers based on a segmentation of the domain. This operator supports a second domain-directed axis of exploration. The segmentation is user-specified and application dependent. The CSP peel operator results in the CSP of the bivariate field restricted to an input segment of the domain. 

\myparagraph{Segmentation.} 
A molecular subgroup is represented by a subset of atoms that constitute the molecule. Following previous work~\cite{masood2021visual}, we compute a weighted Voronoi tessellation~\cite{Aurenhammer1987powerdiag} in order to partition the molecule into atomic regions. The input point set to the tessellation algorithm consists of atom locations and radii. The volume occupied by each molecular subgroup is now represented as a union of atomic regions. The volumetric regions corresponding to each subgroup define a segmentation of the molecule and constitute the input to the CSP peel operator. Other geometric or topology-based segmentation may be used if deemed appropriate. Atoms that constitute the various subgroups of interest are assumed to be available as input to the visualization pipeline.

\myparagraph{Interpretation.} 
Interpretation of a peeled CSP is driven by the specific application and the set of queries being posed. In the electronic transitions application, the primary goal is to analyze the donor or acceptor behavior of each subgroup under consideration. If the peeled CSP aligns with the X-axis or specifically consists of the region satisfying $| \phi_h | > | \phi_p |$, then we classify the subgroup as a donor. A horizontal line $\phi_p = 0$ implies that the particle NTO equals zero within the corresponding region in the domain and there is no charge gain. Rather, this region has lost charge because the hole NTO   {$\phi_h \neq 0$}. Similarly, if the CSP aligns with the Y-axis, then the corresponding region in the domain represents an acceptor. In \autoref{fig:teaser} (right), Thiophene is classified as donor while Quinoxaline as an acceptor.

A large difference between the area covered by the CSPs of subgroups may help infer which subgroup is a stronger donor or acceptor. Alternatively, a consistent CSP for a subgroup across different molecules indicates a unique property of the subgroup. For example, we compare the behavior of copper within different complexes in \autoref{fig:cu-ligands}. Peeling also helps identify the nature of the electronic transition. Local Excitation (LE) refers to the scenario where both hole and particle NTOs are located within a subgroup, and Charge Transfer Excitation (CT) where the NTOs are located within different subgroups. 
LE may be identified by comparing CSPs of individual subgroups. If the contribution from all subgroups except one is small then we may conclude that the electronic transition is local to a subgroup. For example, we identify State~10 as a LE state in Cu-PHE-PHEOME, see \autoref{fig:cu-lect}. 

\subsection{Joint application of CSP lens and peel operators}
The CSP lens and peel operators may be applied independently (single axis) or simultaneously (both axes) to explore the bivariate field depending on application requirements. The CSP peel operator could function as a second operator following the CSP lens operator, as shown in \autoref{fig:csp_lens-peel}. The Y-axis shows the peeled CSPs for each subgroup. Queries along the X-axis may be posed independently either on the CSP corresponding to the entire molecule or for each subgroup. Such queries help identify donor and acceptor regions lying within a subgroup. For example, we are able to segment the donor and acceptor regions within the Quinoxaline subgroup. {The donor CSP restricted to the Quinoxaline subgroup is highlighted (red box). \autoref{fig:lens-illustration} shows how to highlight the donor region using fiber surfaces extracted from this CSP. The green fiber surfaces are restricted to the Quinoxaline subgroup as compared to \autoref{fig:tq-120-lens}, which is generated using only the lens operator.}

\subsection{User input} To generate the matrix shown in \autoref{fig:csp_lens-peel}, a user specifies the lens operators (X-axis, columns) and domain segmentation (Y-axis, rows). The user may select a particular CSP for further analysis as shown in \autoref{fig:lens-illustration}. The selection of control polygon can be manual or automated if it can be defined mathematically. For this application, donor strength is a key property under study and derives other characteristics like local excitation and charge transfer. It can be set and changed manually to generate the corresponding control polygons and fiber surfaces. {Paraview~\cite{Ayachit2015paraview} along} with TTK~\cite{Tierny2018ttk} are used for interaction and to generate all the visualizations presented in this paper.

\begin{figure*}[!t]
    \centering
    \begin{tabular}{c@{\hskip3pt}c@{\hskip1pt}c@{\hskip1pt}c@{\hskip1pt}c@{\hskip1pt}c@{\hskip5pt}c}
    & $0^{\circ}$ & $60^{\circ}$ & $90^{\circ}$ & $120^{\circ}$ & $180^{\circ}$ & Relaxed ($35.3^{\circ}$)
    \\
    \cmidrule(lr){2-6} \cmidrule(lr){7-7}
    \raisebox{0.2\height}{\rotatebox{90}{Geometry}}
    &
    \includegraphics[width=0.15\textwidth]{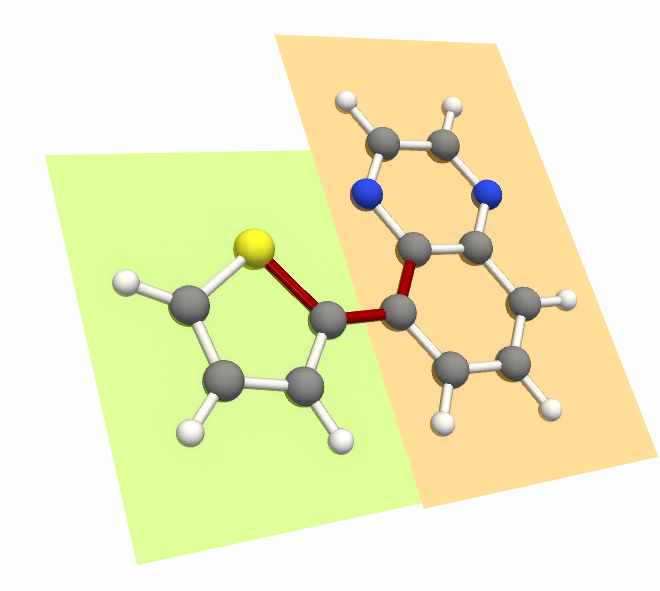}
    &
    \includegraphics[width=0.15\textwidth]{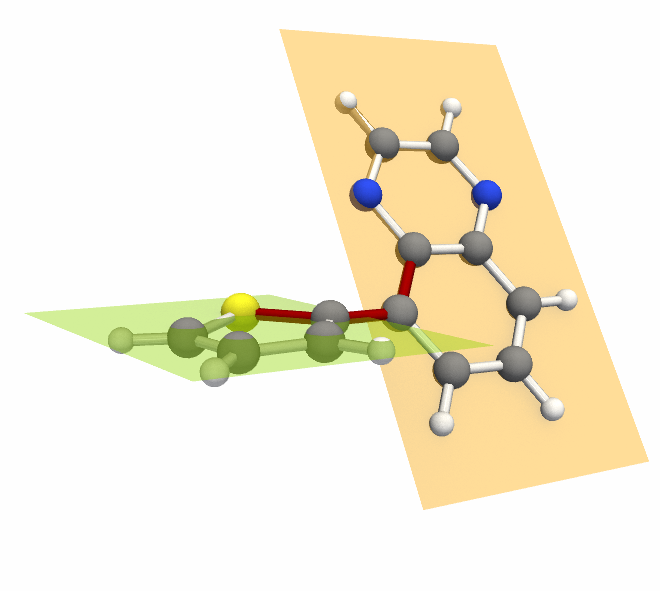}
    &
    \includegraphics[width=0.15\textwidth]{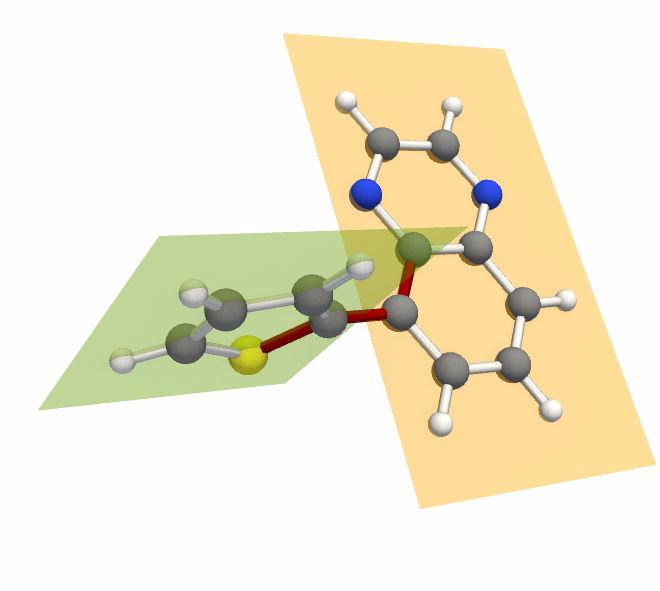}
    &
    \includegraphics[width=0.15\textwidth]{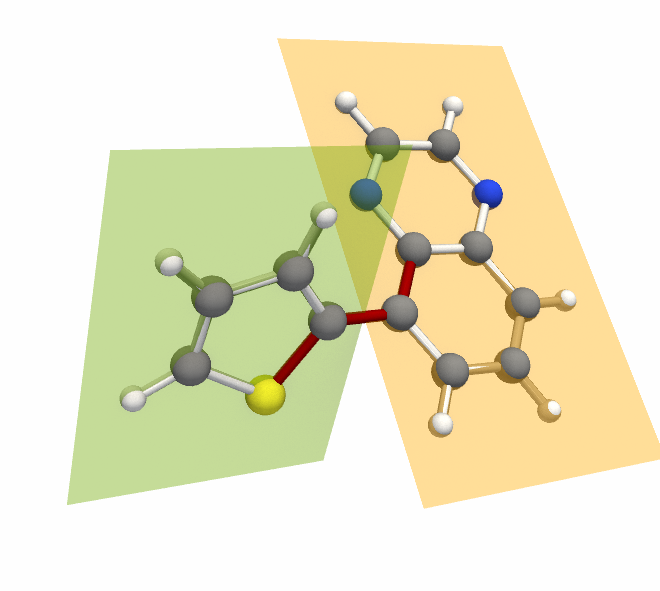}
    &
    \includegraphics[width=0.15\textwidth]{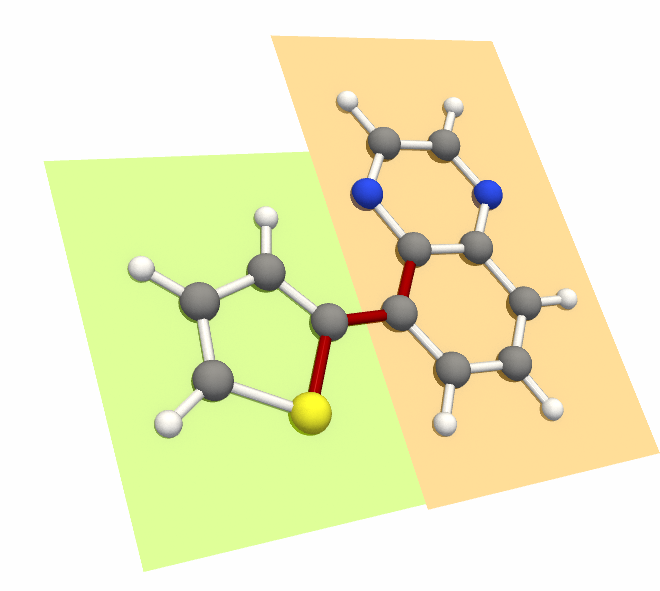}
    &
    \includegraphics[width=0.15\textwidth]{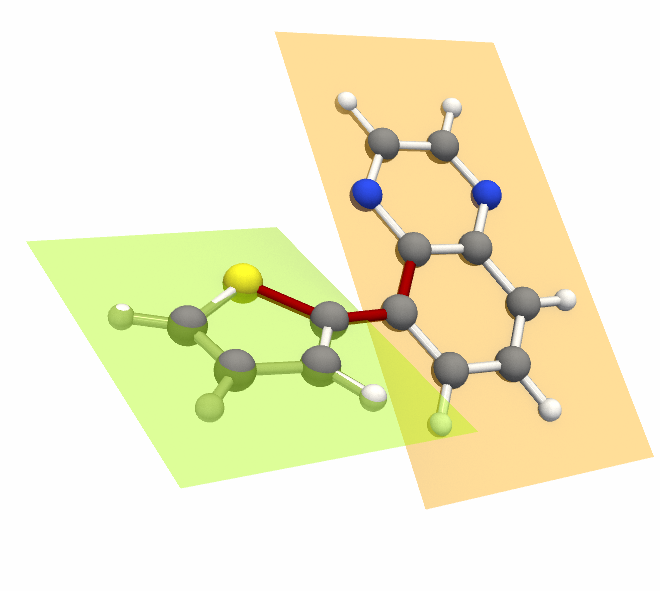}
    \\
    \raisebox{0.9\height}{\rotatebox{90}{Th-Qu}}
    &
    \includegraphics[width=0.15\textwidth]{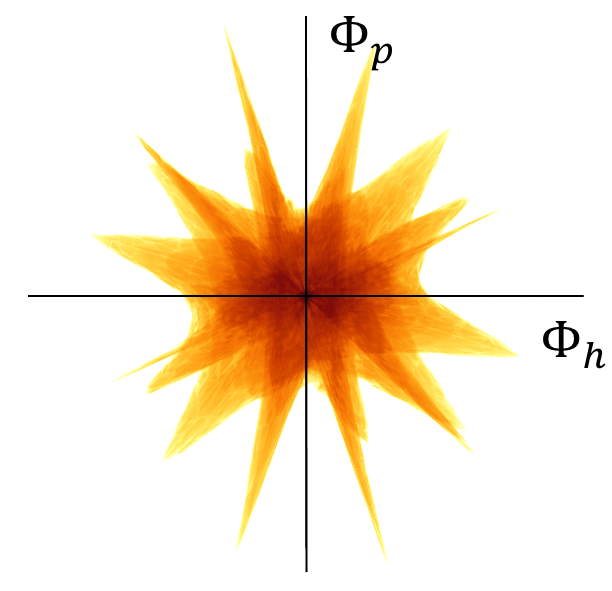}
    &
    \includegraphics[width=0.15\textwidth]{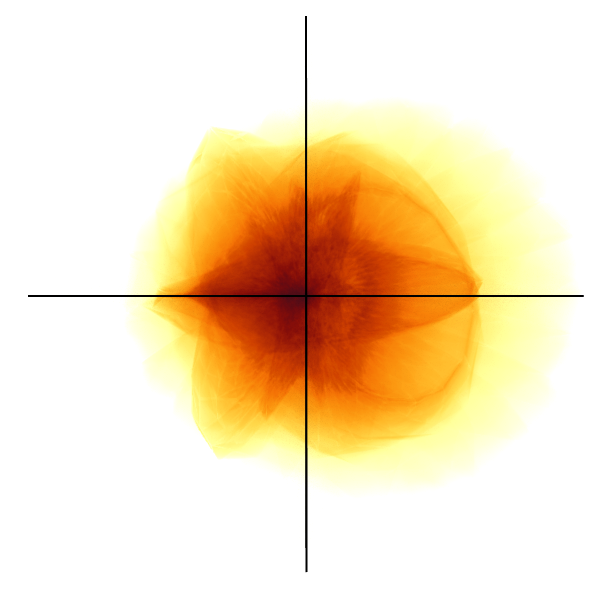}
    &
    \includegraphics[width=0.15\textwidth]{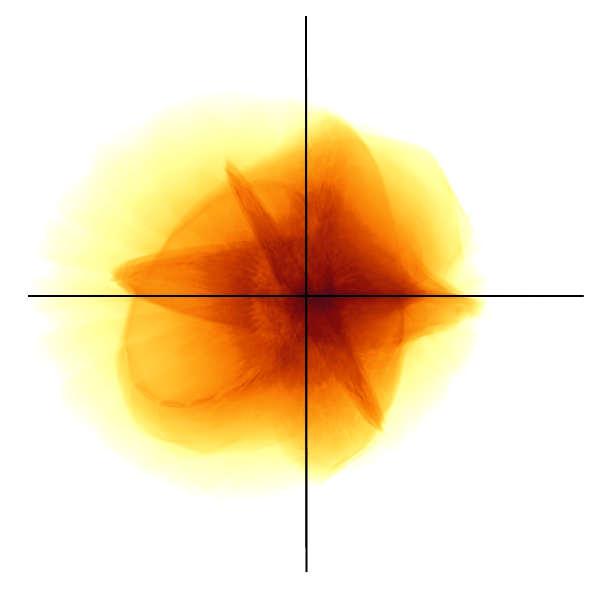}
    &
    \includegraphics[width=0.15\textwidth]{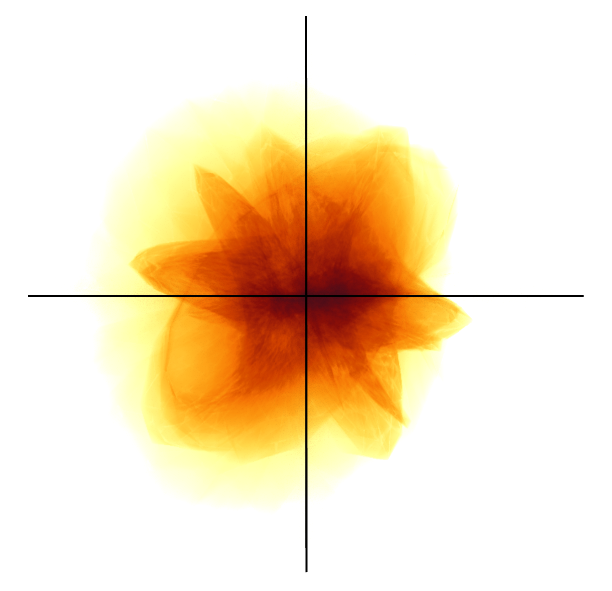}
    &
    \includegraphics[width=0.15\textwidth]{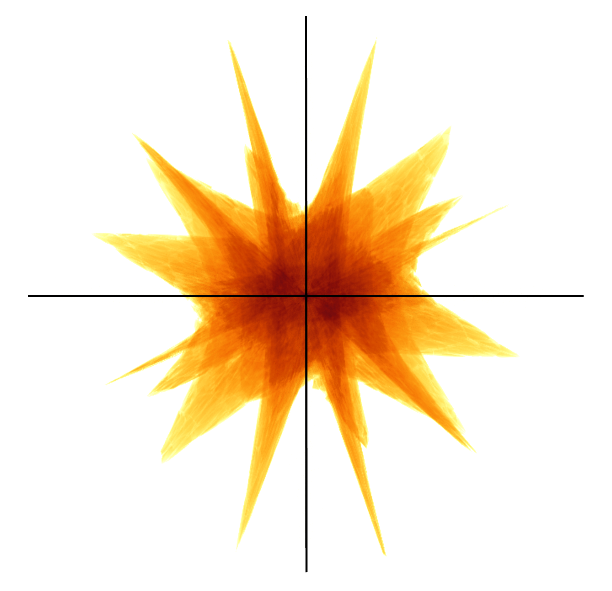}
    &
    \includegraphics[width=0.15\textwidth]{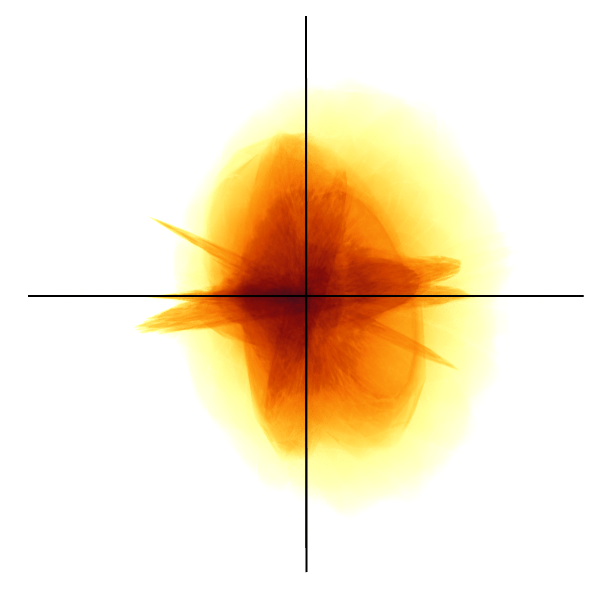}
    \\
    \raisebox{0.1\height}{\rotatebox{90}{Thiophene (Th)}}
    &
    \includegraphics[width=0.15\textwidth]{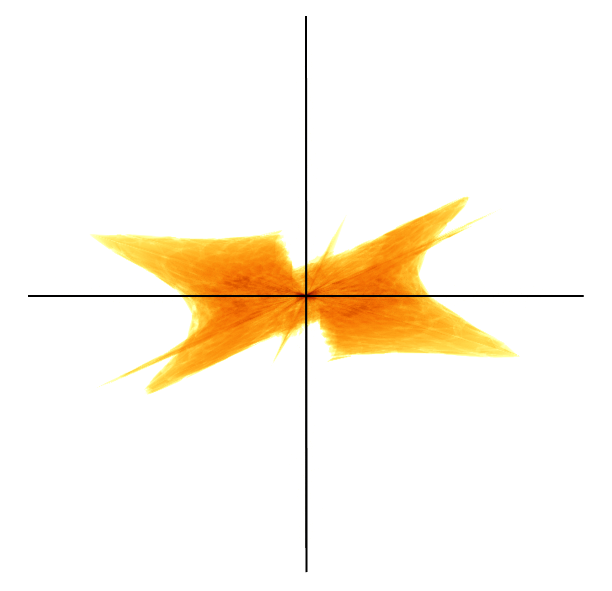}
    &
    \includegraphics[width=0.15\textwidth]{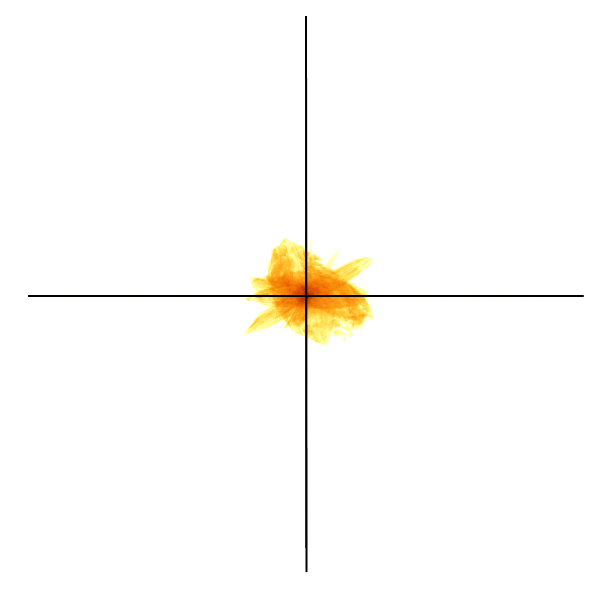}
    &
    \includegraphics[width=0.15\textwidth]{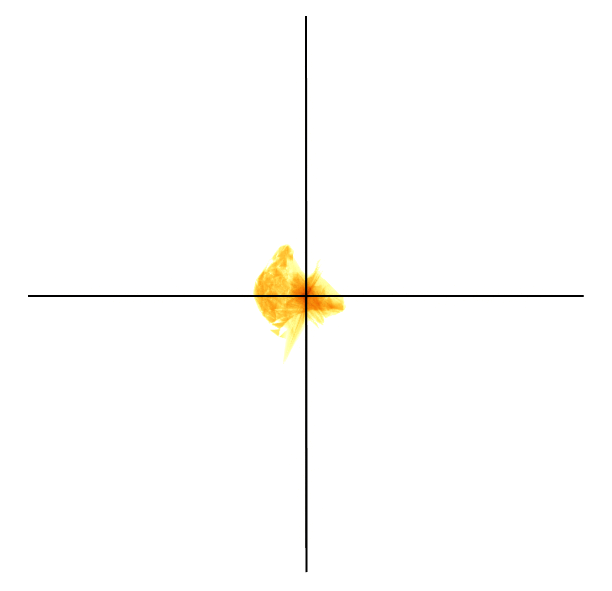}
    &
    \includegraphics[width=0.15\textwidth]{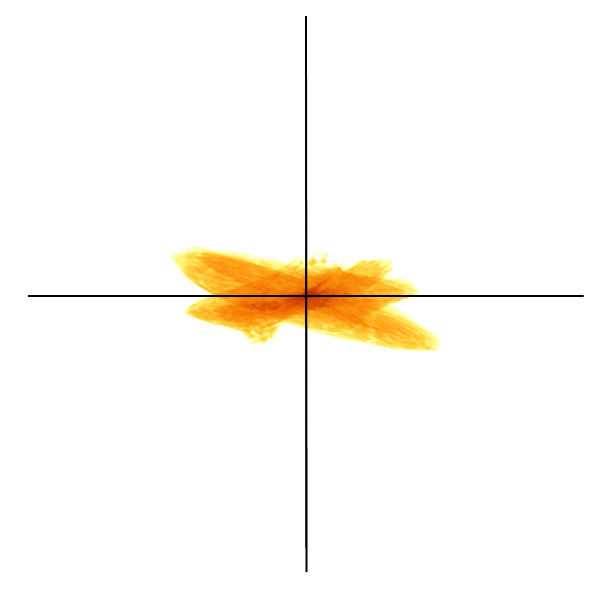}
    &
    \includegraphics[width=0.15\textwidth]{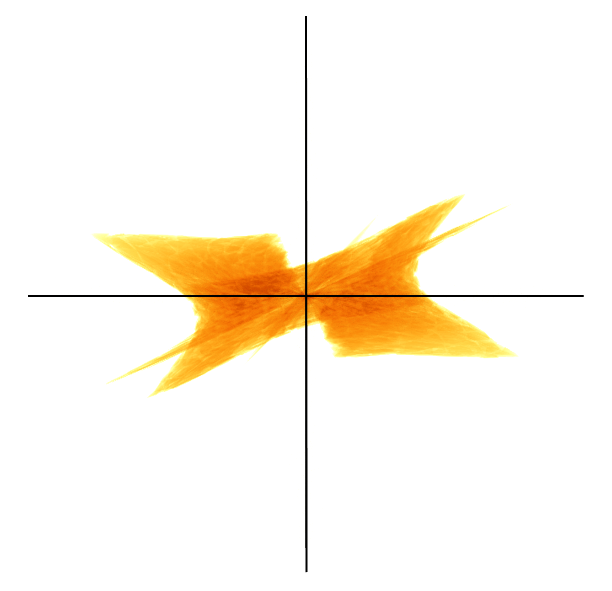}
    &
    \includegraphics[width=0.15\textwidth]{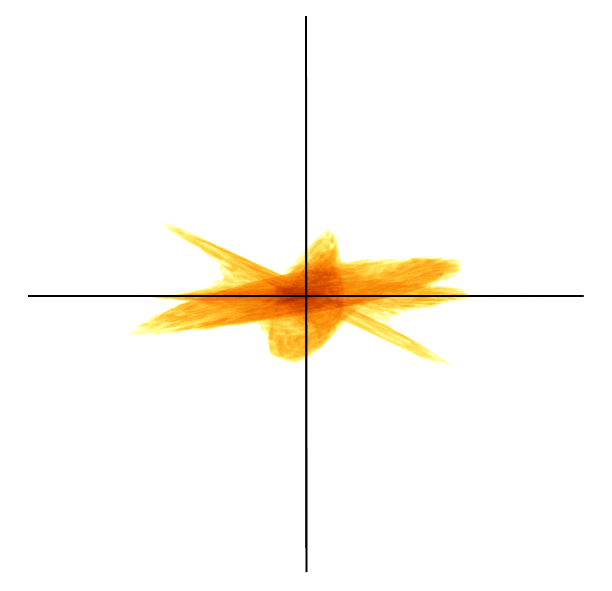}
    \\
    & $\Delta = 0.440$ & $\Delta = 0.026$ & $\Delta = -0.006$ & $\Delta = 0.227$ & $\Delta = 0.429$ & $\Delta = 0.480$
    \\
    \raisebox{0.04\height}{\rotatebox{90}{Quinoxaline (Qu)}}
    &
    \includegraphics[width=0.15\textwidth]{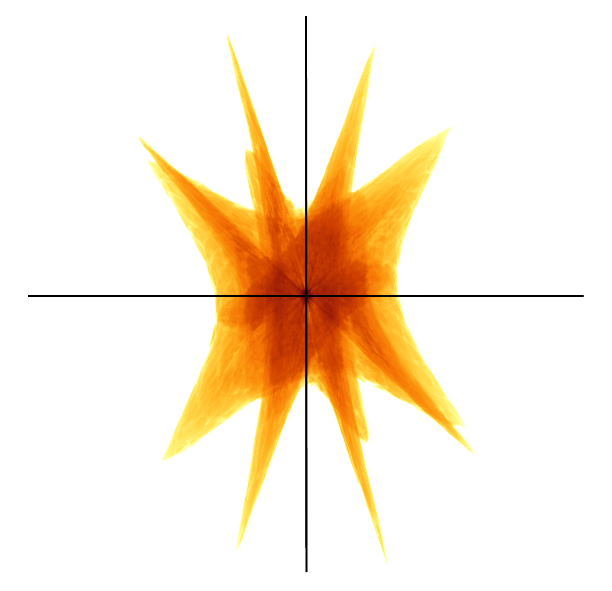}
    &
    \includegraphics[width=0.15\textwidth]{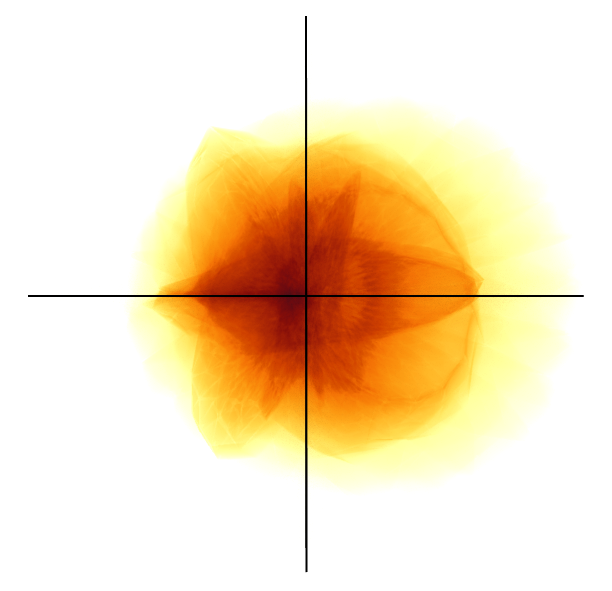}
    &
    \includegraphics[width=0.15\textwidth]{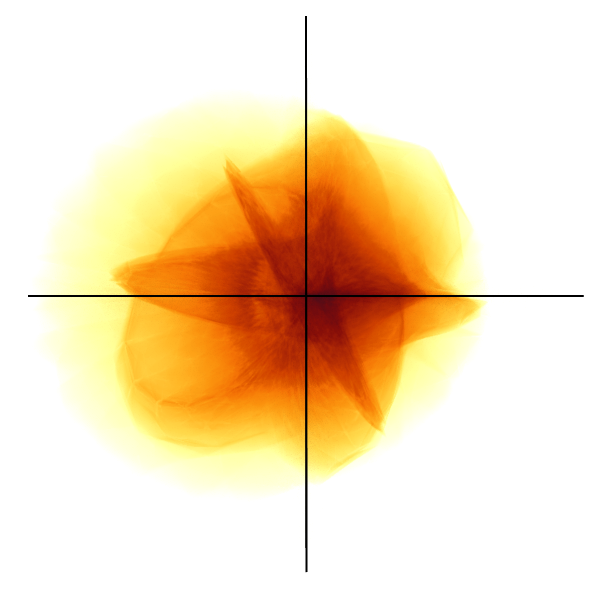}
    &
    \includegraphics[width=0.15\textwidth]{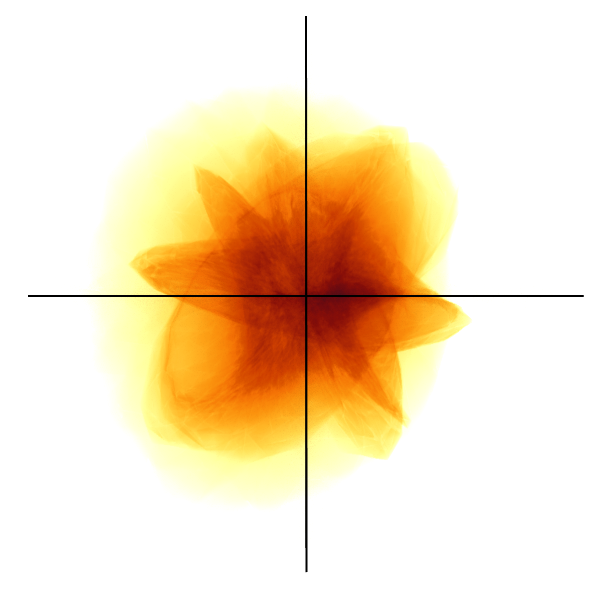}
    &
    \includegraphics[width=0.15\textwidth]{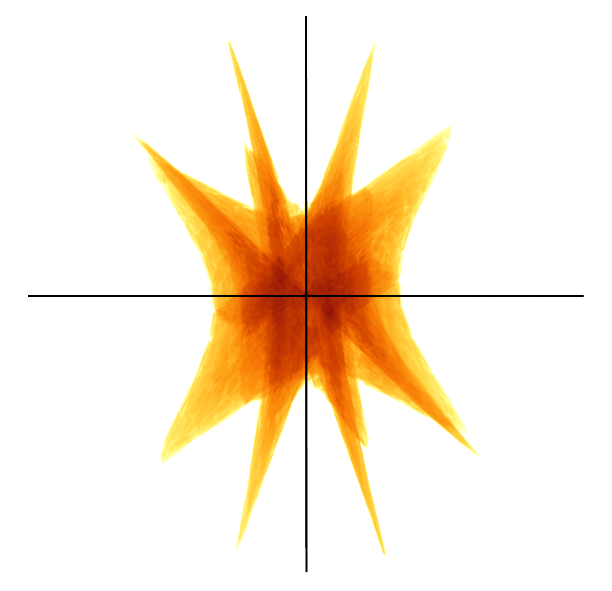}
    &
    \includegraphics[width=0.15\textwidth]{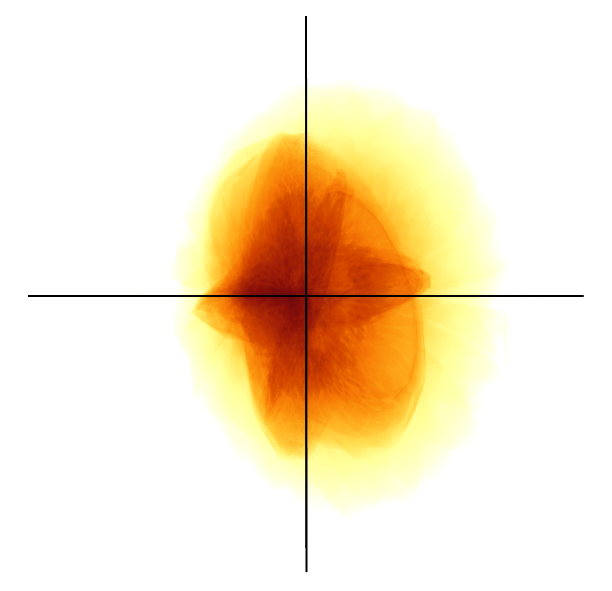}
    \\
    & $\Delta = -0.440$ & $\Delta = -0.026$ & $\Delta = 0.006$ & $\Delta = -0.227$ & $\Delta = -0.429$ & $\Delta = -0.480$
    \end{tabular}
    \caption{CSP peel operator helps in the analysis of TQ (green plane - Thiophene, orange plane - Quinoxaline) for varying dihedral angle based conformations of state~1. In the $60^{\circ}$ and $90^{\circ}$ conformations, the peeled CSP for Thiophene covers a small region suggesting local excitation within Quinoxaline. $90^{\circ}$ exhibits least charge transfer from Thiophene to Quinoxaline. Thiophene behaves as a donor within other molecular conformations.
    } 
    \label{fig:tq-geometry}
\end{figure*}

\subsection{Quantification}
\label{sec:quantification}
The CSP lens operator facilitates range space queries and the CSP peel operator facilitates domain-based queries as shown in \autoref{fig:teaser}. The use of both operators relies on subsequent visual inspection of the results. In some cases, a visual inspection may not suffice for a comprehensive analysis. In \autoref{fig:cu-ligands}, a visual observation of the horizontally aligned CSPs is sufficient to infer that copper behaves as a donor in the three molecules, but we are unable to determine the order of their donor strengths. We present quantitative measures that help address this problem. Each point $i$ in the CSP is assigned a weight $w_i$,  and accumulated to obtain a measure $\Delta$ for the CSP,  
\begin{align*}
\Delta = \sum_{i=1}^{N} d_i \cdot w_i
\end{align*}
Here, $d_i$ represents the point density and $N$ is the total number of points that depends on the resolution of the CSP. The weight $w_i$ is assigned based on the specific property that we wish to quantify. The point density $d_i$ determines the contribution of a point $i$ towards the property. $\Delta$ may also be viewed as an integral over all fibers in the domain using Federer's Co-Area Formula~\cite{morgan2000geometric,scheidegger2008revisiting}.

The properties that we want to quantify are closely related to the characteristics captured by the CSP lens operators. A constant weight $w_i = 1$ represents the identity lens, and the resulting value of $\Delta$ is equal to the volume of the spatial domain. For electronic transitions, a weight $w_i = |\phi_{h_i}|^2$ represents the mask function of the hole charge lens operator and the corresponding $\Delta$ is equal to 
the hole charge of the subgroup. In order to compute the `donor strength' of a particular subgroup, we subtract the total charge gained by the subgroup from the charge it lost. This quantity is equal to the difference between the mask functions of the donor and acceptor lens operators, $w_i = |\phi_{h_i}|^2 - |\phi_{p_i}|^2$. The donor strength is now defined as 
\begin{align*}
\Delta = \sum_{i=1}^{N} d_i \cdot (|\phi_{h_i}|^2 - |\phi_{p_i}|^2)
\end{align*}
In the rest of this paper, the term $\Delta$ always refers to donor strength.

\myparagraph{Computation.} The CSP may be computed using \textsc{ttk}. The accuracy of the CSP computed using this implementation is dependent on its image resolution. Similarly, the resolution of the CSP determines the accuracy of the quantification. 
A finer resolution representation of the CSP results in increased accuracy of $\Delta$ but at the cost of a computationally intensive procedure to generate the high resolution CSP.

Due to above mentioned issues, we compute the donor strength using a 2D discrete histogram based implementation. All histograms are reported on a $1000 \times 1000$ grid whose x-axis represents $\phi_h$ and y-axis represents $\phi_p$. A vertex from the spatial domain is mapped to the  appropriate bin in the histogram, and its function value is rounded to the midpoint of the bin. This rounding causes an approximation error in the computation of the quantitative measure. We observe in our experiments that the error in the donor strength due to midpoint approximation is in the range ($\pm 0.001, \pm 0.032$). A detailed description of the method used for computing the donor strength and tables listing the errors are available in the supplementary material.
%
%
%
%
%
%

\section{Results}\label{sec:results}
We now present results of two case studies where the two operators reveal the donor-acceptor behavior, the variation of the behavior with different geometric conformations of a molecule and with varying ligands in a family of molecular complexes. The molecular complexes considered in the case studies are well studied. Insights from visual analysis help confirm some of their known properties. The electronic transitions are calculated using the Gaussian software package~\cite{Frisch2016Gaussian} and the hole and particle NTOs are generated with a script developed in-house. The CSPs and fiber surfaces are computed using \textsc{ttk}~\cite{Tierny2018ttk}. The method development, case study identification, and interpretations  were conducted in close collaboration with a theoretical chemist. 

\subsection{Case Study 1: Thiophene-Quinoxaline}
The first case study investigates different geometric conformations of TQ using the CSP peel operator, as shown in the first row of \autoref{fig:tq-geometry}. Our analysis focuses on the first excited state of the molecule.
Thiophene (the 5-member ring) is generally considered to be a donor subgroup but, as our study shows, this property varies depending on the dihedral angle between the subgroups. The $\pi$-bond conjugation (delocalization of electrons over the molecule)  can be broken (at 90°) which results in different types of excitation, local vs. charge transfer, for different angles. The CSP analysis results are shown in \autoref{fig:tq-geometry}.

The $\pi$-bond conjugation is maximum in $0^{\circ}$ and $180^{\circ}$ as the orbitals are fully delocalized over the molecule. The CSPs for these two angles are near identical, and hence capture the similarity between the two conformations. The peeled CSPs corresponding to Thiophene for $0^{\circ}$ and $180^{\circ}$ are aligned along the X-axis implying that Thiophene is the donor subgroup, the Quinoxaline CSPs are more aligned along the Y-axis suggesting that it is an acceptor in both conformations.
The area covered by Thiophene CSP shrinks towards origin from $0^{\circ}$ to $60^{\circ}$ and reduces to a minimum at $90^{\circ}$. This suggests that both hole and particle NTOs for Thiophene reduce and hence the charge transfer from Thiophene to Quinoxaline also decreases as the dihedral angle increases. 
Such a behavior is expected because the $\pi$-bond conjugation is minimum at $90^{\circ}$. The area increases when the dihedral angle increases further to $120^{\circ}$.
Quinoxaline CSP for $60^{\circ}$ and $90^{\circ}$ is similar to the CSP of the molecule, while Thiophene CSP covers a small region near the origin. This suggests local excitation in these conformations, primarily within Quinoxaline. The behavior changes towards charge transfer excitation again at $120^{\circ}$.
In the relaxed geometry, where the molecule is at an energy optimal dihedral angle ($35.3^{\circ}$), we observe that Thiophene acts as a donor, which is again the expected behavior.

The subgroup CSPs are annotated with a quantity that represents donor strength of the corresponding segment. We observe that Thiophene and Quinoxaline have similar donor strengths for $0^{\circ}$ and $180^{\circ}$. The value of $\Delta$ computed for Thiophene supports relative ordering of donor strength in different geometric configurations, 
\begin{align*}
{0^{\circ} \approx 180^{\circ} > 120^{\circ} > 60^{\circ} > 90^{\circ}.}
\end{align*}
It is generally expected that charge transfer is favored by a higher conjugation but the behavior also depends on exact orbital alignment. We note that $\Delta_{Th(35.3^\circ)} > \Delta_{Th(0^\circ)}$. It is difficult to conclude just by visual inspection that donor strength is the highest in relaxed geometry. 

The quantification of donor strength is useful for the chemist to compare the charge transferred in a precise manner. The $\Delta$ value complements the information obtained via visual inspection and vice-versa. The $\Delta$ value helps determine whether the subgroup is behaving as donor or acceptor overall but visual analysis provides further insights. In this case study, calculating only the $\Delta$ values does not help conclude which of the two subgroups undergoes local excitation in the $60^{\circ}$ and $90^{\circ}$ conformations. This is because the $\Delta$ values are small for both the subgroups involved in the electronic transition. However, the CSPs help resolve this ambiguity. The Quinoxaline's CSP is more spread than that of Thiophene, revealing that the excitation is happening within the former group.
For the relaxed geometry, by visual analysis, one may incorrectly assign a local excitation character for this transition since the CSP for Quinoxaline looks similar to the CSP of the whole molecule and Thiophene CSP's contribution could be considered small. Although, the $\Delta$ values show that the charge donated by Thiophene to Quinoxaline is large enough and comparable to $0^{\circ}$ and $180^{\circ}$ conformations. Hence, with the help of quantification, we can conclude that it is clearly a charge transfer scenario.
For the conformation at $120^{\circ}$, we observe a $\Delta$ value indicating charge transfer from Thiophene to Quinoxaline. However, the CSP for Thiophene mostly extends along the donor axis while the CSP of Quinoxaline extends on both the axes, indicating that this transition must be a mixture of local excitation on Quinoxaline (\autoref{fig:tq-120-lens}) and charge transfer from Thiophene to Quinoxaline.
It is already known from the domain expert that planar conformation ($0^{\circ}$ and $180^{\circ}$) will favor the delocalization of $\pi$ electron over the full molecule and therefore favor charge transfer between the two subgroups. In this spirit, it is not surprising to observe that the $90^{\circ}$ conformation present no charge transfer ($\Delta$). Interestingly, the relaxed geometry which is not planar is the one presenting the largest $\Delta$ but close to that of $0^{\circ}$ and $180^{\circ}$ conformations, indicating that the planarity cannot be the only criterion to explain the amount of charge transfer observed.

\begin{figure}[!ht]
    \centering
    \begin{tabular}{c@{\hskip3pt}c@{\hskip1pt}c@{\hskip1pt}c@{\hskip1pt}c@{\hskip1pt}}
    & \small{Cu-PHE-PHE} & \small{Cu-PHE-PHEOME} & \small{Cu-PHE-XANT}
    \\
    \cmidrule(lr){2-4}
    &
    \includegraphics[width=0.14\textwidth]{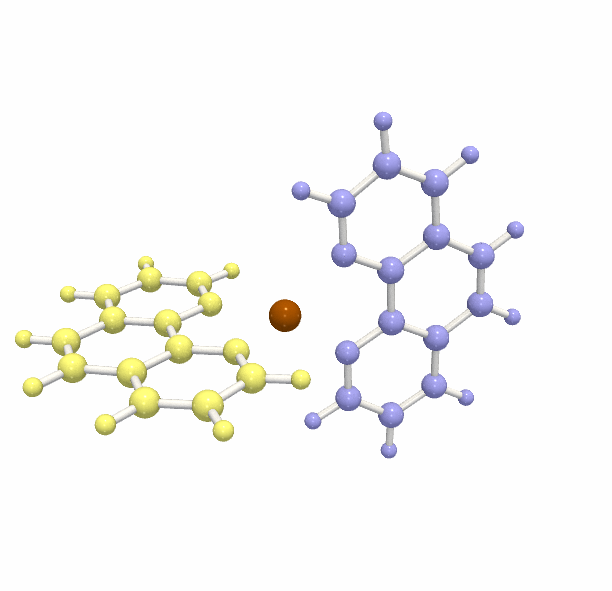}
    &
    \includegraphics[width=0.14\textwidth]{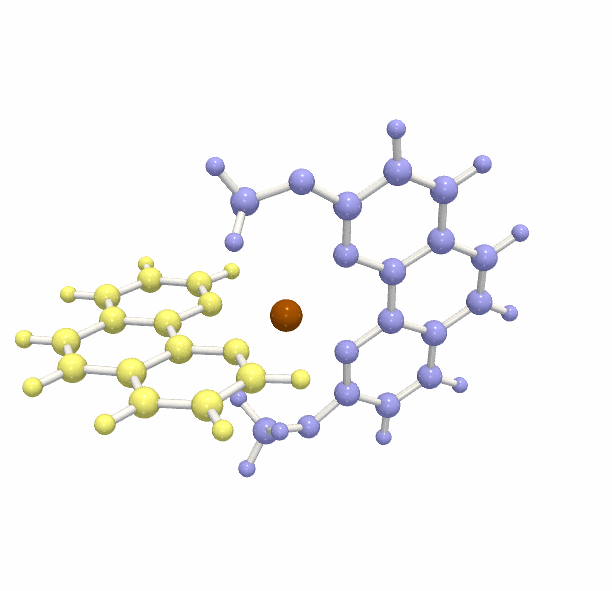}
    &
    \includegraphics[width=0.14\textwidth]{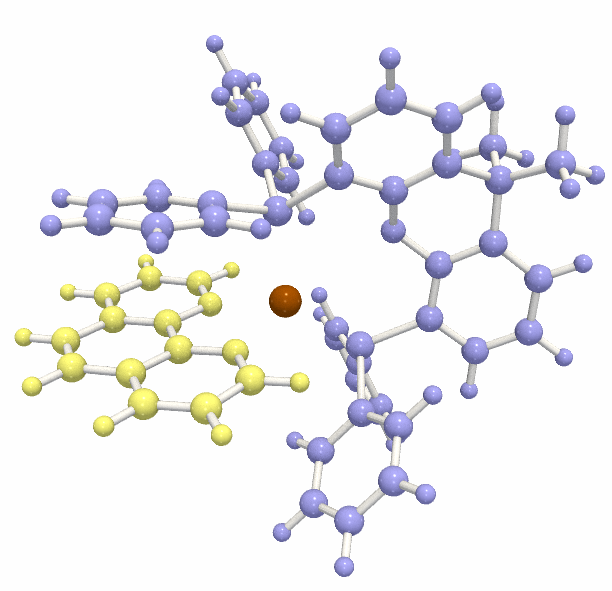}
    \\
    \raisebox{0.05\height}{\rotatebox{90}{\small{Cu-PHE-Ligand}}}
    &
    \includegraphics[width=0.14\textwidth]{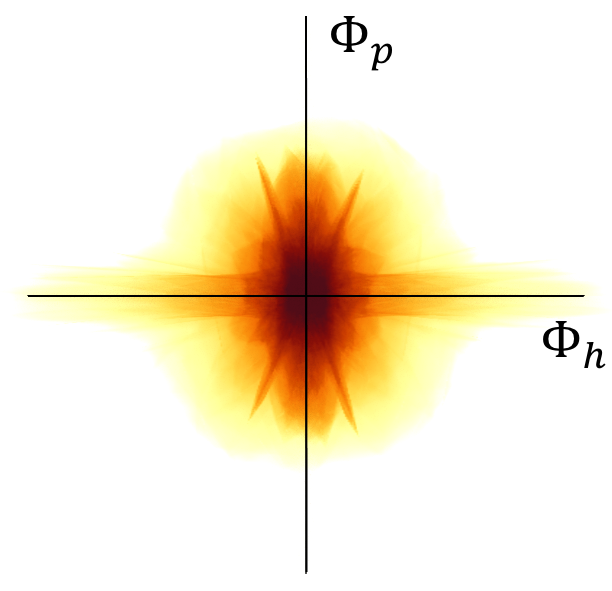}
    &
    \includegraphics[width=0.14\textwidth]{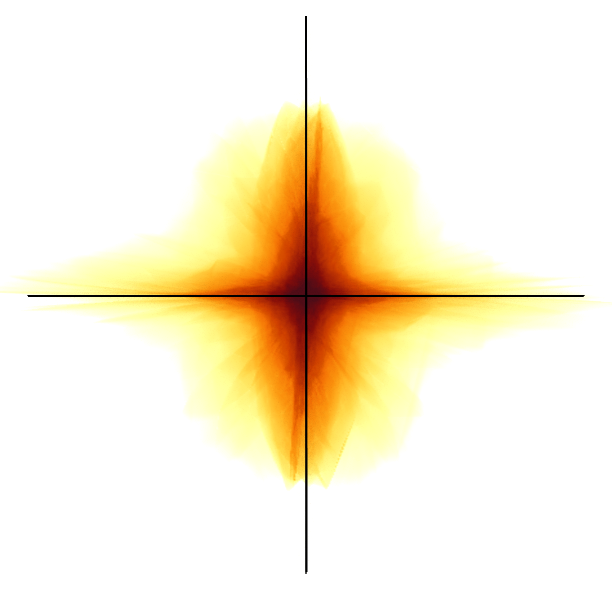}
    &
    \includegraphics[width=0.14\textwidth]{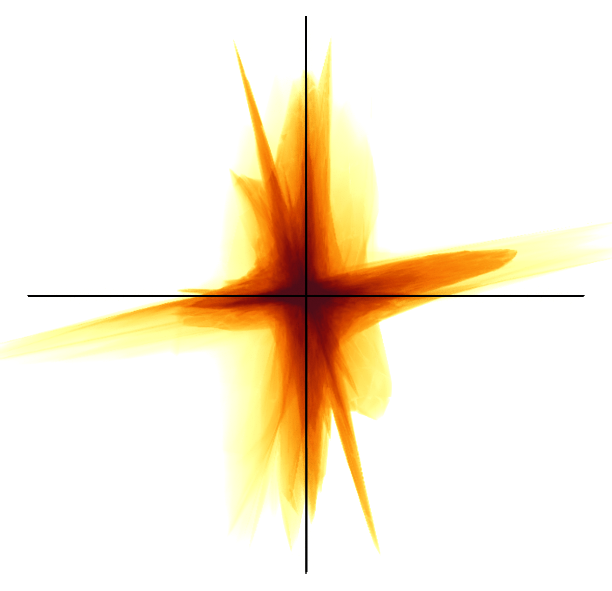}
    \\
    \raisebox{2.3\height}{\rotatebox{90}{\small{Cu}}}
    &
    \includegraphics[width=0.14\textwidth]{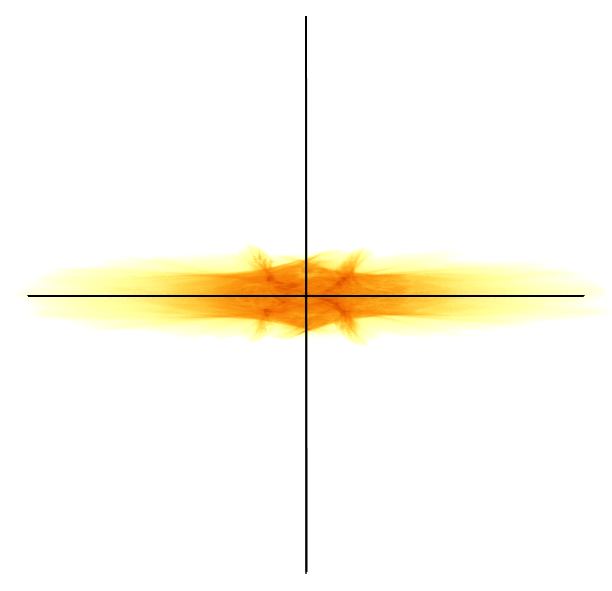}
    &
    \includegraphics[width=0.14\textwidth]{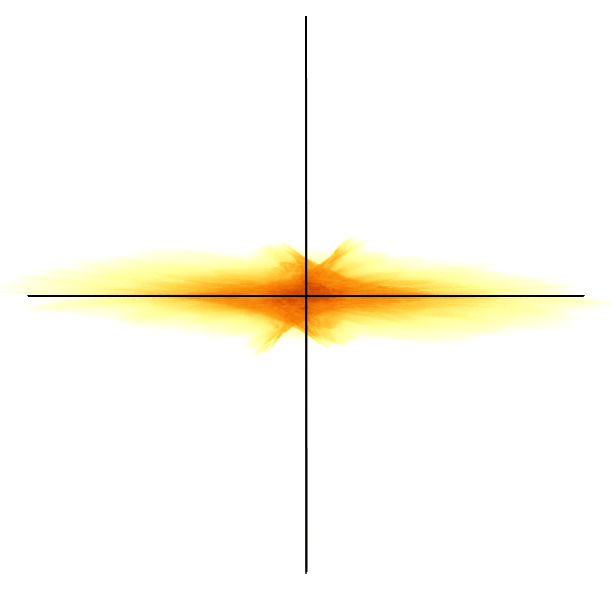}
    &
    \includegraphics[width=0.14\textwidth]{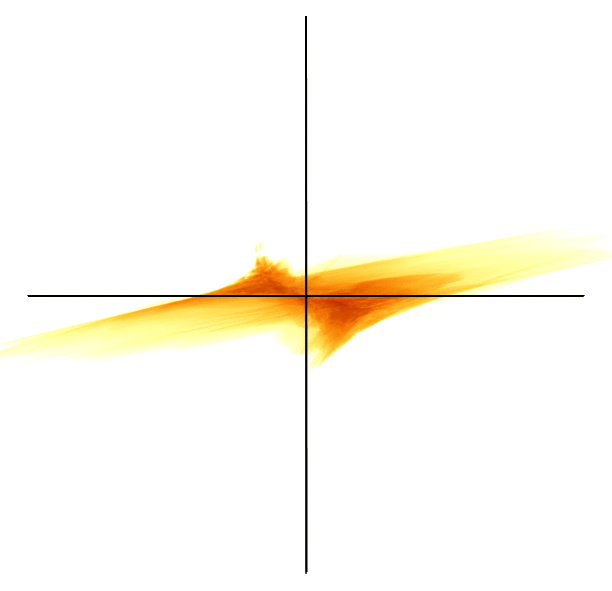}
    \\
    & \footnotesize{$\Delta = 0.681$} & \footnotesize{$\Delta = 0.634$} & \footnotesize{$\Delta = 0.328$}
    \\
    \raisebox{1.4\height}{\rotatebox{90}{\small{PHE}}}
    &
    \includegraphics[width=0.14\textwidth]{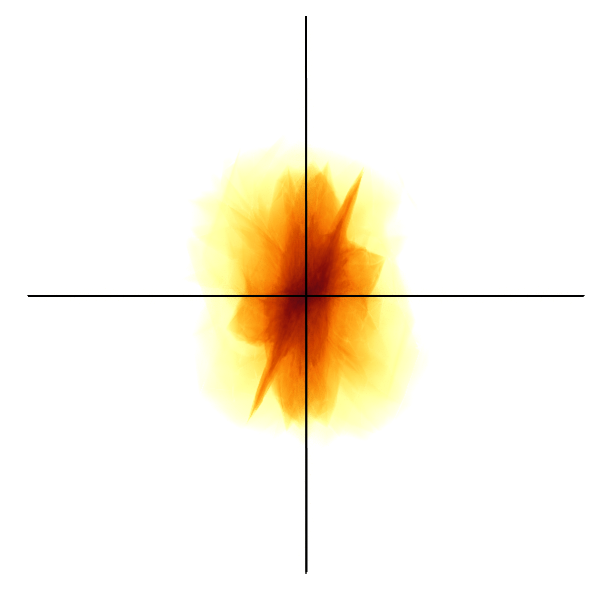}
    &
    \includegraphics[width=0.14\textwidth]{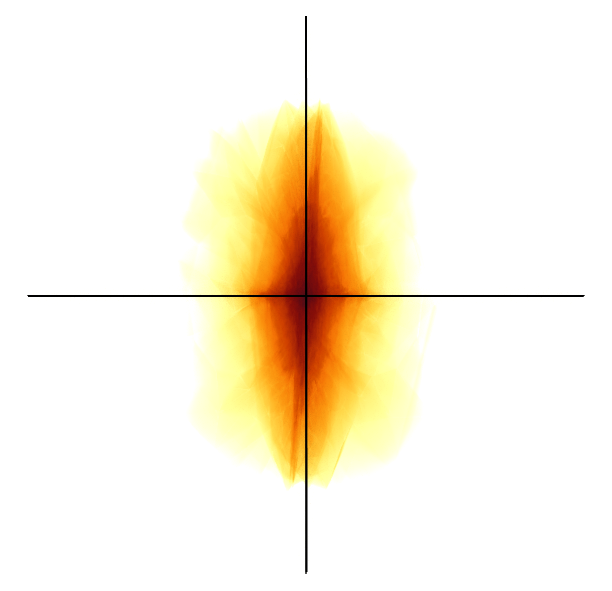}
    &
    \includegraphics[width=0.14\textwidth]{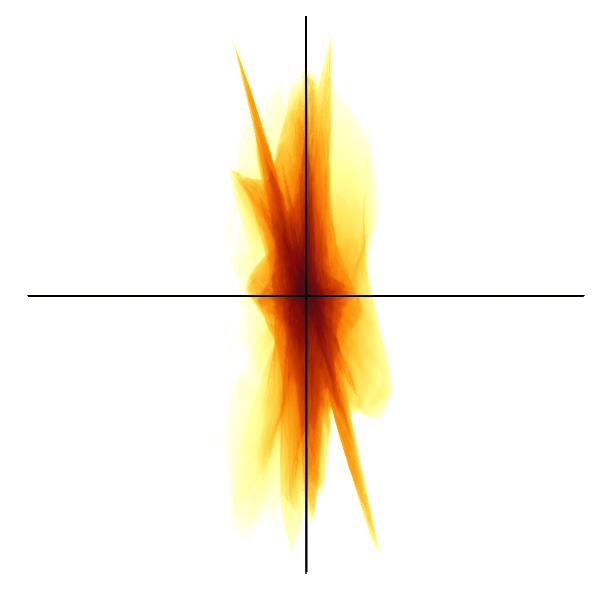}
    \\
    & \footnotesize{$\Delta = -0.289$} & \footnotesize{$\Delta = -0.811$} & \footnotesize{$\Delta = -0.876$}
    \\
    \raisebox{0.7\height}{\rotatebox{90}{\small{Ligand}}}
    &
    \includegraphics[width=0.14\textwidth]{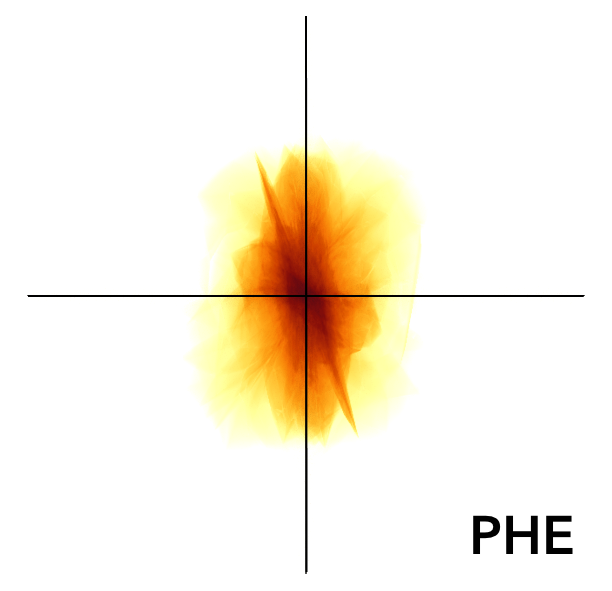}
    &
    \includegraphics[width=0.14\textwidth]{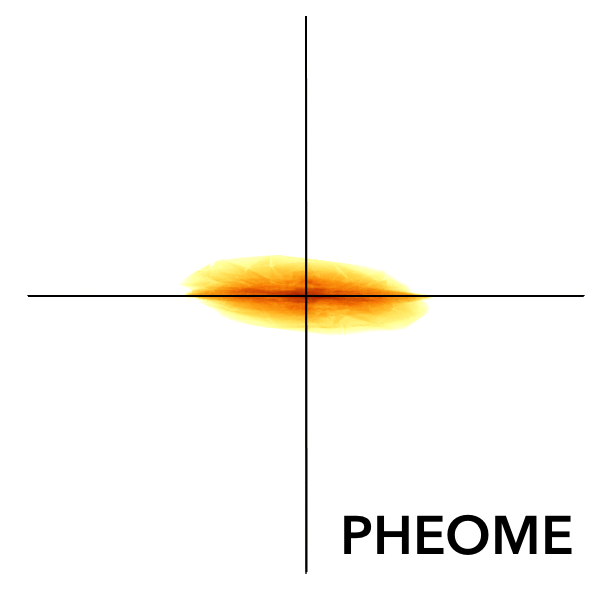}
    &
    \includegraphics[width=0.14\textwidth]{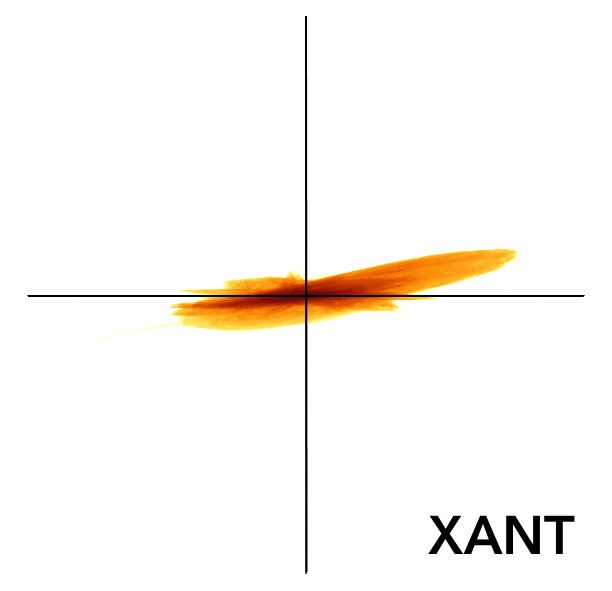}
    \\
    & \footnotesize{$\Delta = -0.391$} & \footnotesize{$\Delta = 0.177$} & \footnotesize{$\Delta = 0.547$}
    \\
   \end{tabular}
    \caption{CSP peel operator helps analyze copper complexes with different ligand configurations, all in state~1. Cu behaves as a donor and PHE behaves as an acceptor within all complexes. The charge transfer from Cu is symmetric in Cu-PHE-PHE.} 
    \label{fig:cu-ligands}
\end{figure}
\begin{figure}[!ht]
    \centering
    \begin{tabular}{c@{\hskip3pt}c@{\hskip1pt}c@{\hskip1pt}c@{\hskip1pt}c@{\hskip1pt}cc@{\hskip1pt}c}
    & \multicolumn{2}{c}{\small{Cu-PHE-PHEOME}} &\multicolumn{2}{c}{\small{Cu-PHE-XANT}} \\
    \cmidrule(lr){2-3} \cmidrule(lr){4-5}
    & \small{State 9} & \small{State 10} & \small{State 3} & \small{State 10}
    \\
    \cmidrule(lr){2-3} \cmidrule(lr){4-5}
     \raisebox{0.00001\height}{\rotatebox{90}{\small{Cu-PHE-Ligand}}}
    &
    \includegraphics[width=0.11\textwidth]{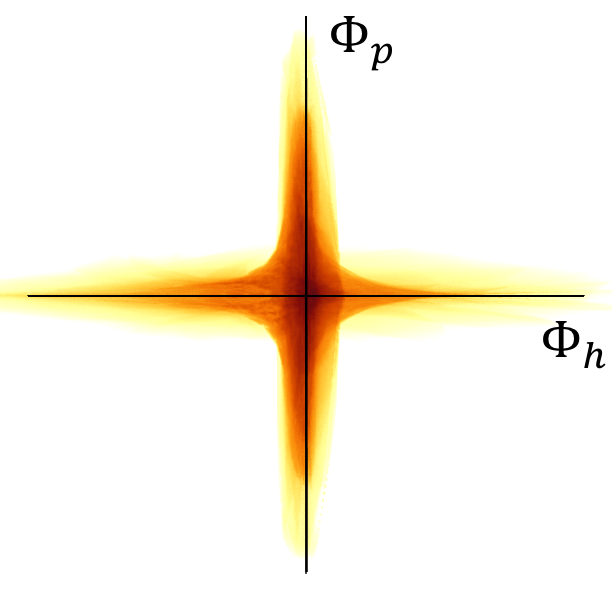}
    &
    \includegraphics[width=0.11\textwidth]{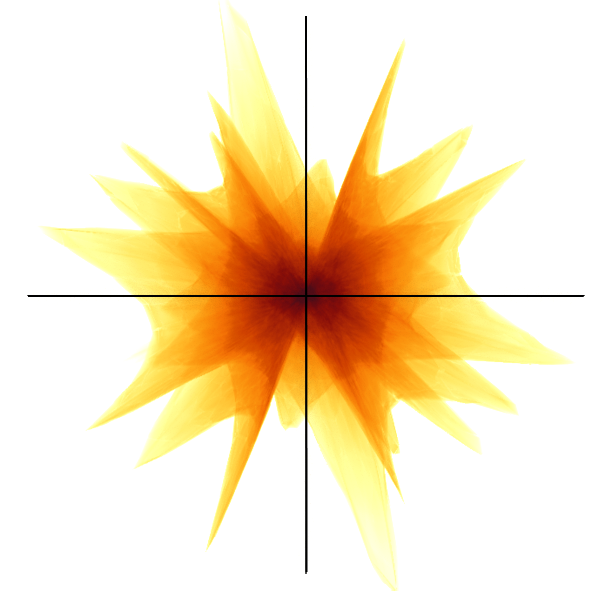}
    &
    \includegraphics[width=0.11\textwidth]{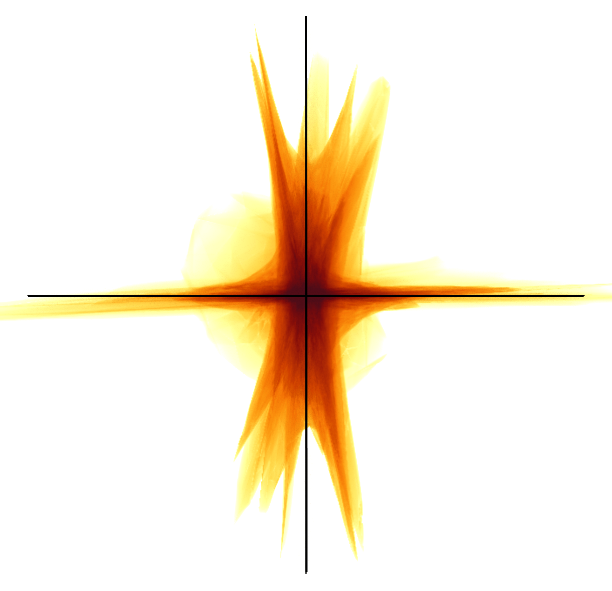}
    &
    \includegraphics[width=0.11\textwidth]{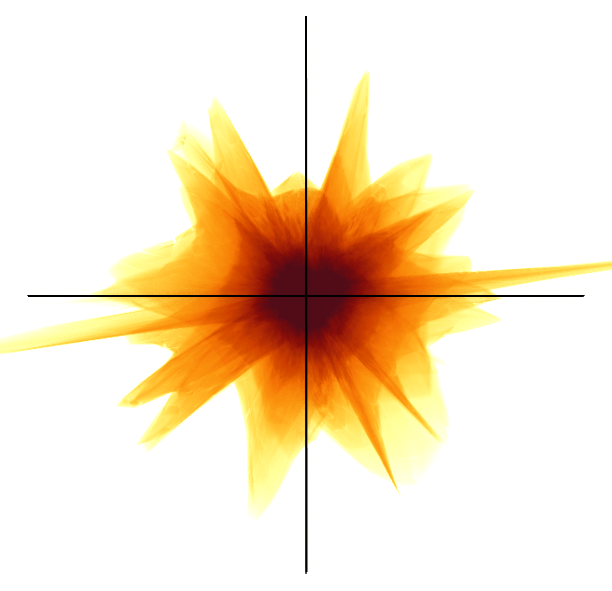}
    \\
    \raisebox{1.9\height}{\rotatebox{90}{\small{Cu}}}
    &
    \includegraphics[width=0.11\textwidth]{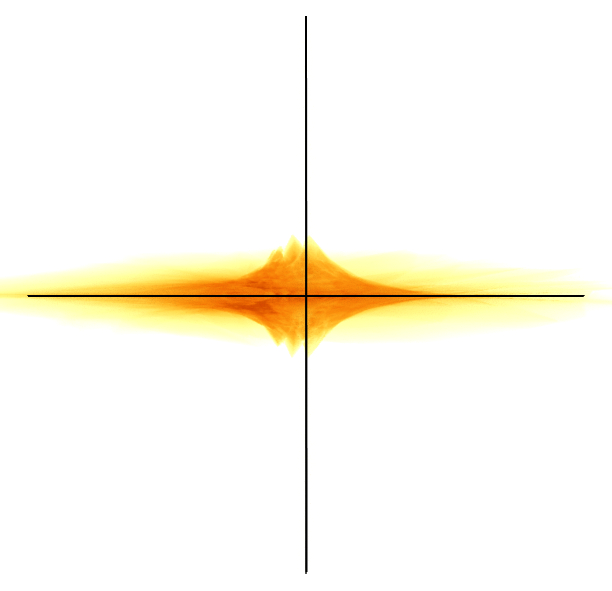}
    &
    \includegraphics[width=0.11\textwidth]{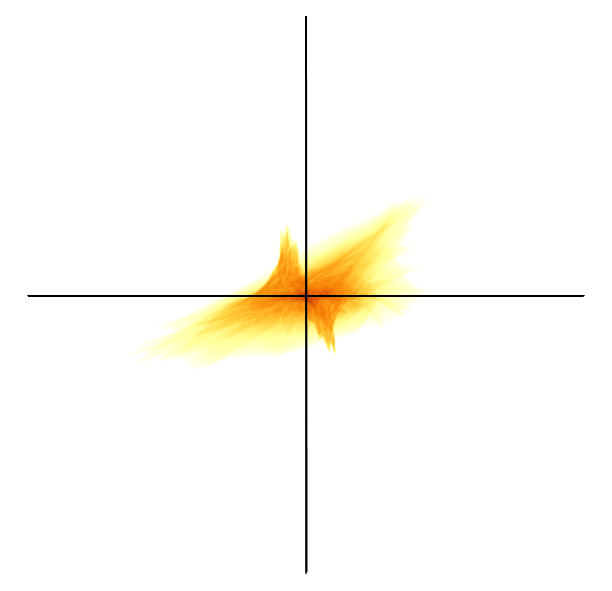}
    &
    \includegraphics[width=0.11\textwidth]{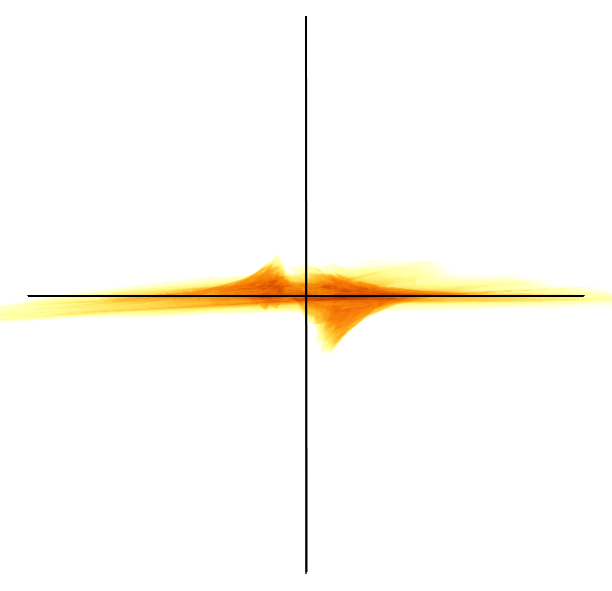}
     &
    \includegraphics[width=0.11\textwidth]{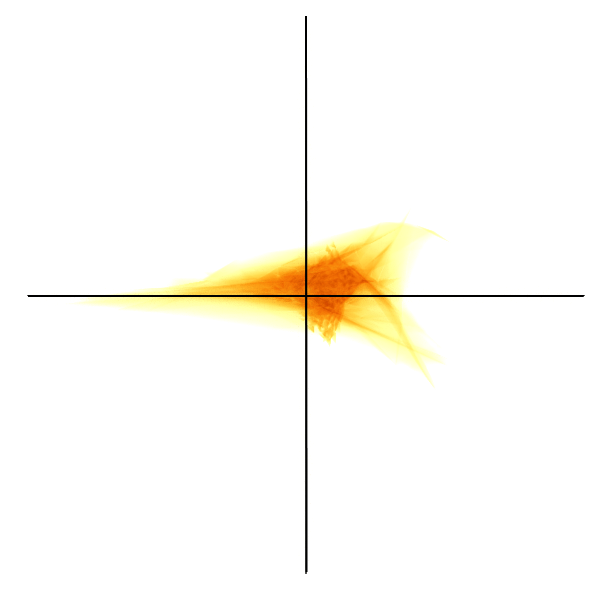}
    \\
    & \footnotesize{$\Delta = 0.933$} & \footnotesize{$\Delta = 0.033$} & \footnotesize{$\Delta = 0.335$} & \footnotesize{$\Delta = 0.031$}
    \\
    \raisebox{1\height}{\rotatebox{90}{\small{PHE}}}
    &
    \includegraphics[width=0.11\textwidth]{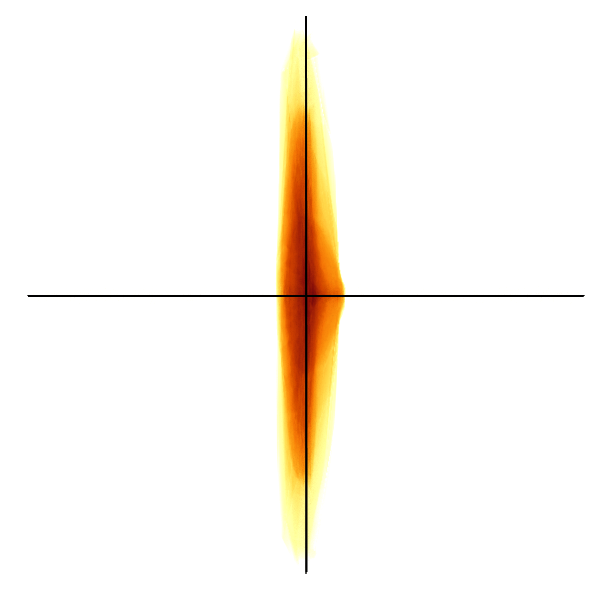}
    &
    \includegraphics[width=0.11\textwidth]{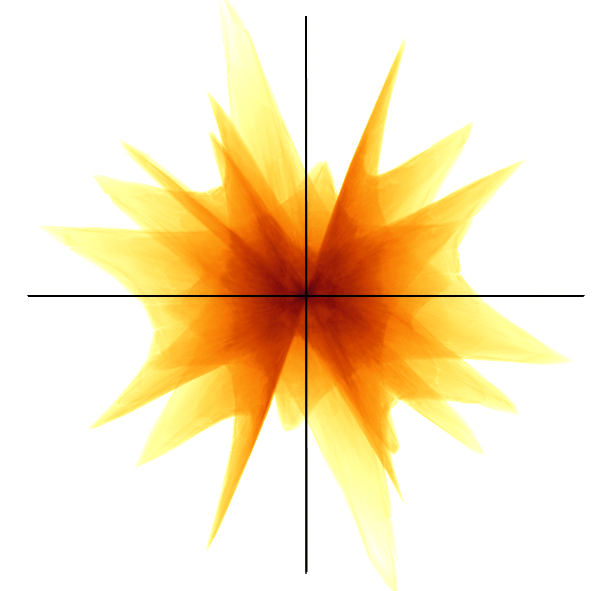}
    &
    \includegraphics[width=0.11\textwidth]{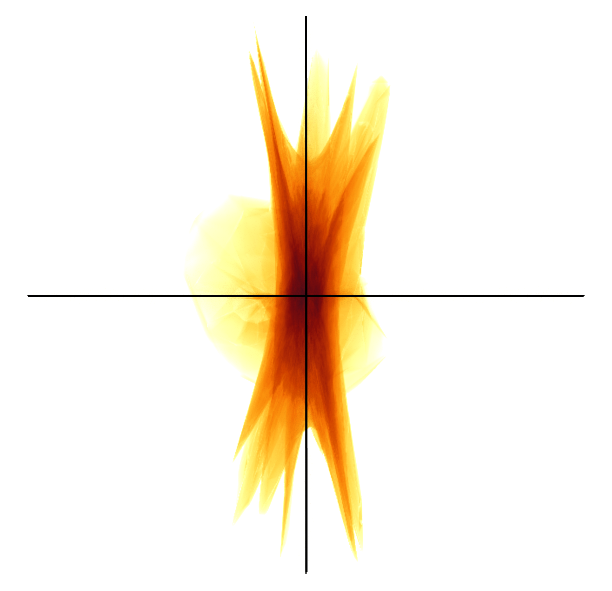}
    &
    \includegraphics[width=0.11\textwidth]{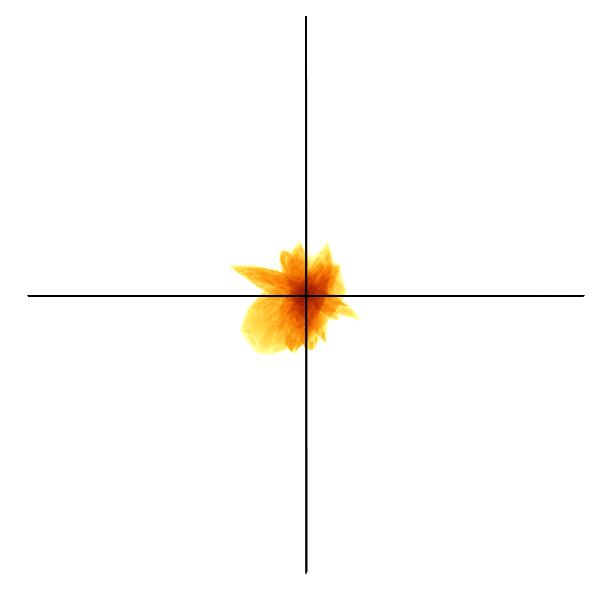}
    \\
    & \footnotesize{$\Delta = -0.943$} & \footnotesize{$\Delta = -0.025$} & \footnotesize{$\Delta = -0.906$} & \footnotesize{$\Delta = -0.022$}
    \\
    \raisebox{0.4\height}{\rotatebox{90}{\small{Ligand}}}
    &
    \includegraphics[width=0.11\textwidth]{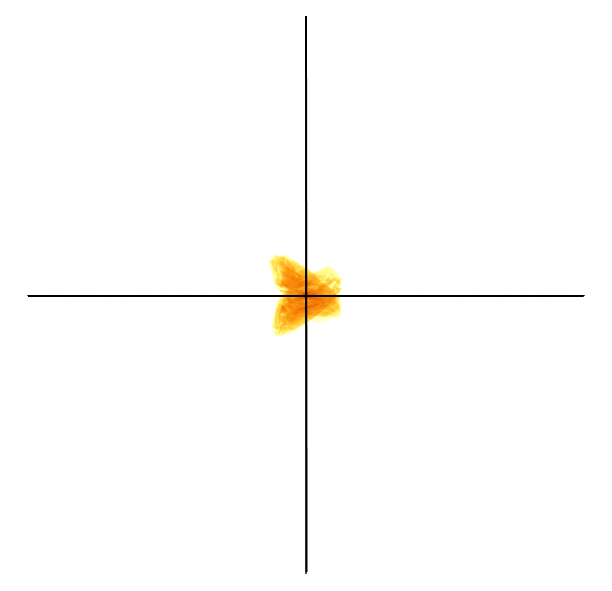}
    &
    \includegraphics[width=0.11\textwidth]{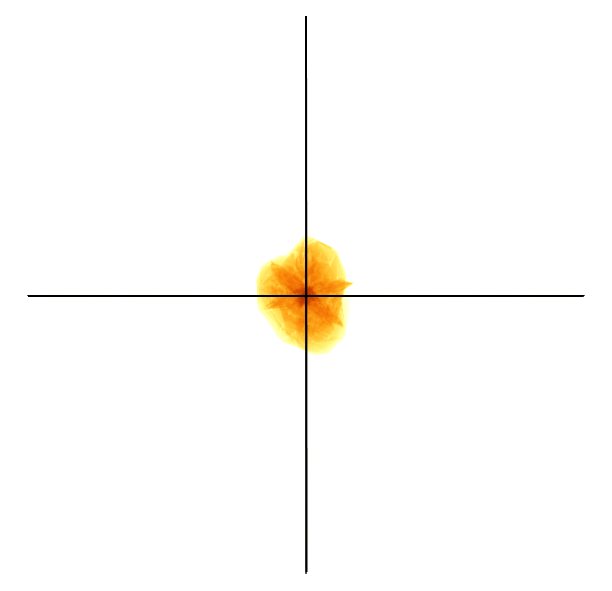}
    &
    \includegraphics[width=0.11\textwidth]{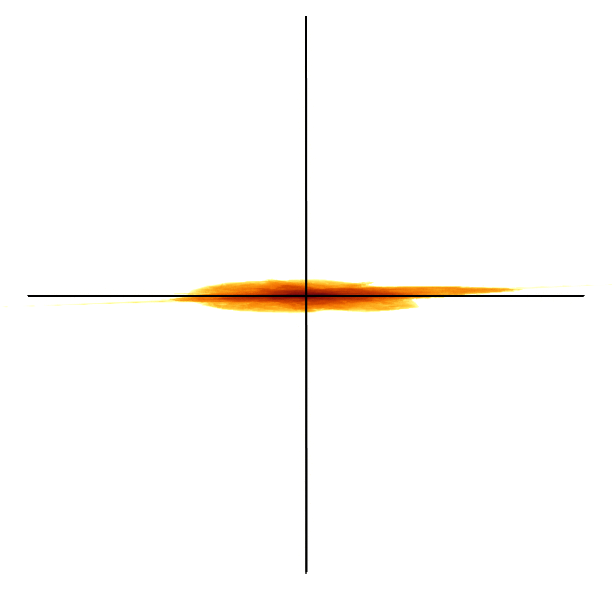}
    &
    \includegraphics[width=0.11\textwidth]{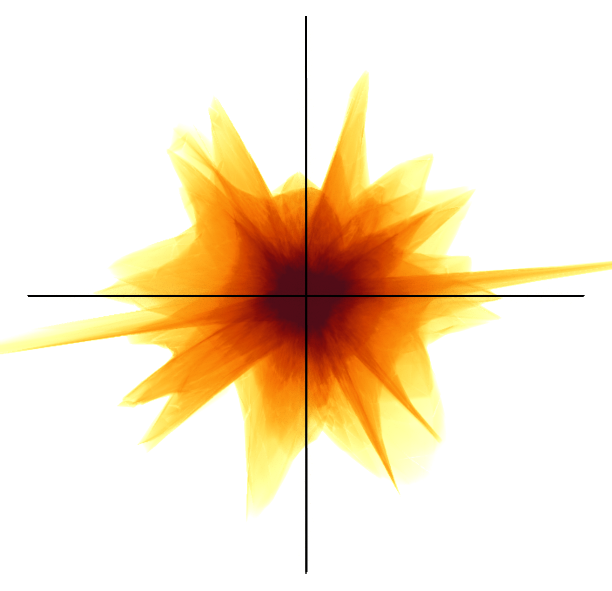}
    \\
    & \footnotesize{$\Delta = 0.010$} & \footnotesize{$\Delta = -0.008$} & \footnotesize{$\Delta = 0.571$} & \footnotesize{$\Delta = -0.009$}
   \end{tabular}
    \caption{CSP peel operator helps compare different excitation states. For Cu-PHE-PHEOME, we observe charge transfer in state~9 and local excitation within PHE in state~10. For Cu-PHE-XANT, in state~3 we note charge transfer excitation and local excitation within XANT in state~10.} 
    \label{fig:cu-lect}
\end{figure}

\subsection{Case Study 2: Copper complexes}
In this case study, we focus on a series of copper complexes used for luminescence applications. The complexes consist of a central copper atom surrounded by two ligands. One ligand is common to all complexes (Phenanthroline, PHE) while the other ligand varies: Phenanthroline (PHE), Dimethoxy Phenanthroline (PHEOME), Diphosphine (XANT), Diphenyl Phenanthroline (PHEPHE) and Dimethyl Phenanthroline (PHEME). The character of the electronic transitions in the complexes defines their properties. Identifying the transitions and studying them in detail helps in the design of novel ligand candidates.

\begin{figure*}[!t]
    \centering
    \includegraphics[width=\textwidth]{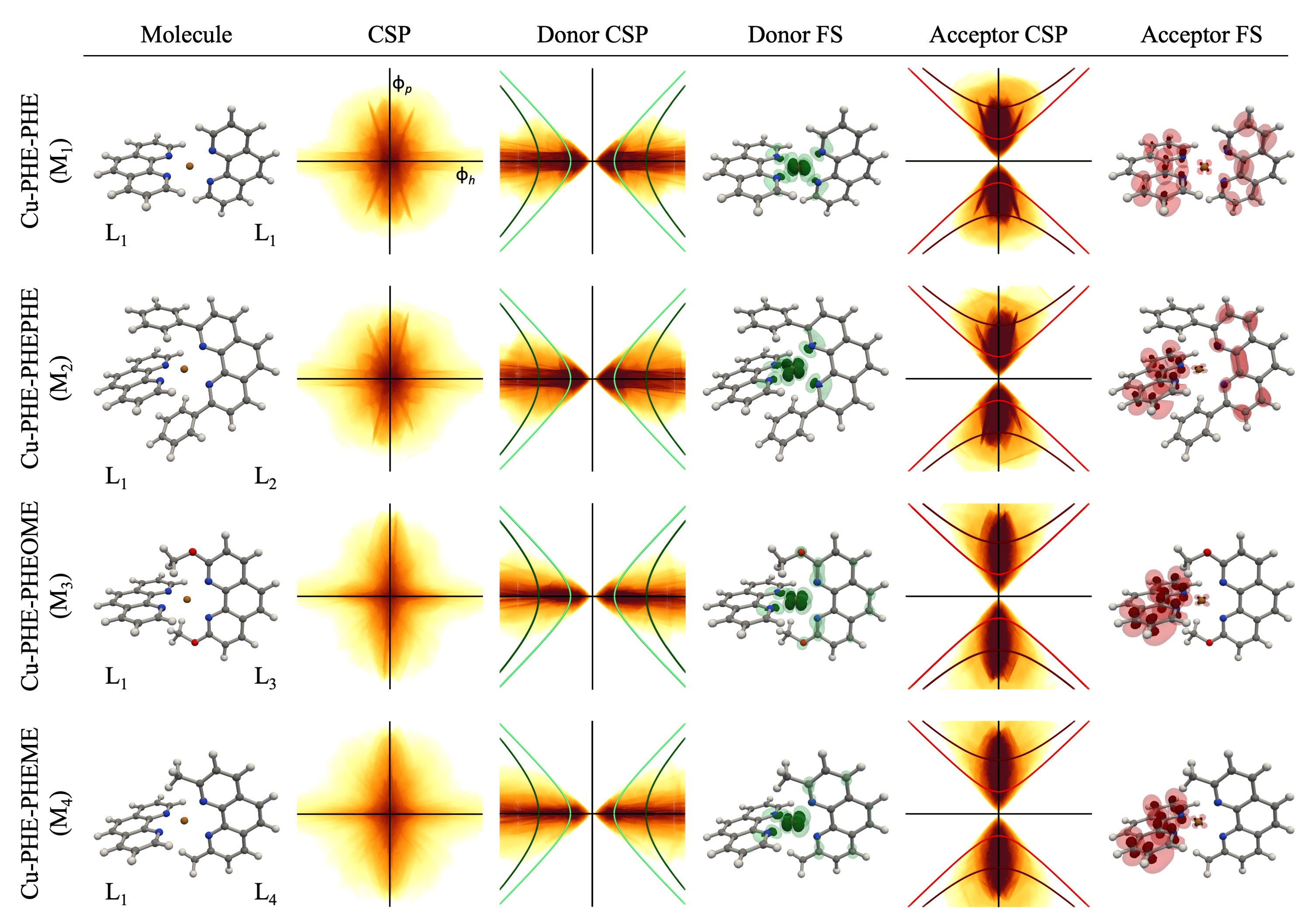}
    \caption{CSP lens operator helps study the change in behavior of $L_1$ after adding substituents, resulting in ligands $L_2, L_3,$ and $L_4$. The light green and light red control polygons represent weaker donor and acceptor behavior compared to their darker counterparts. The red fiber surfaces extend across $L_1$ and $L_2$, which behave like acceptors. The ligand exhibits donor behavior in $M_3$ and $M_4$.}
    \label{fig:substituents-lens}
\end{figure*}

\myparagraph{Varying ligands.}
First, we study the effect of varying the second ligand with a focus on the first excited state of each molecule, see \autoref{fig:cu-ligands}.
Cu behaves as a donor in all configurations, which is the expected behavior. The annotated quantities are also positive for Cu for all conformations \ie $\Delta_{Cu} > 0$.
PHE behaves as an acceptor as indicated by the CSP, which is aligned with the vertical axis, and $\Delta_{PHE} < 0$. An interesting observation is that the charge transfer is expected to be symmetric in Cu-PHE-PHE, from Cu to both PHE ligands. This behavior is observable from the CSPs of the individual PHE ligands.
In Cu-PHE-PHEOME, both Cu and PHEOME are donors. Cu is expected to be a stronger donor. Cu covers a larger span on the X-axis indicating larger values of $\phi_h$ in its atomic region. However, a visual inspection of the CSPs is not sufficient to claim that Cu is indeed the stronger donor. CSPs for Cu and PHEOME appear to be similarly close to origin, but the donor strengths help confirm that Cu is a stronger donor than PHEOME ($\Delta_{Cu} > \Delta_{PHEOME}$).
In Cu-PHE-XANT, both Cu and XANT behave as donors. XANT is expected to be a stronger donor than Cu but a visual inspection is again not sufficient to make such a claim. We observe that $\Delta_{Cu} < \Delta_{XANT}$, which helps confirm the claim. Further, the donor strengths reported here match with previously~reported~values~\cite{masood2021visual}.

\myparagraph{Characterizing an excitation.}
We now compare the behavior of two complexes, Cu-PHE-PHEOME and Cu-PHE-XANT, for different excited states. The aim is to characterize the nature of excitation, local or charge transfer. \autoref{fig:cu-lect} shows the results of the CSP peel operator based visual analysis. 
For Cu-PHE-PHEOME, we observe a charge transfer excitation in state~9. The Cu and PHE CSPs exhibit donor and acceptor behaviors, respectively. We may conclude that a significant charge transfer happens from Cu to PHE. However, in state~10, CSPs of the molecule and that of PHE are similar. Further, contributions from other subgroups appear to be low, thereby indicating local excitation within PHE. The $\Delta$ values match with the visual observations. In state~3 of Cu-PHE-XANT, we observe a charge transfer from Cu and XANT to PHE, and in state~10 a local excitation within XANT. $\Delta$ values again confirm these behaviors. Charge lost by Cu and XANT is gained by PHE in state~3. For state~10, we observe smaller values of $\Delta$ supporting the local excitation behavior but with $\Delta$ value alone, we can not identify which subgroup exhibits local excitation. In this case, it is PHE because the CSP of PHE is more spread along the two axes as compared to the other two subgroups.

\myparagraph{Varying Substituents.}
We now compare behavior of four complexes for the first excited states, Cu-PHE-PHE ($M_1$), Cu-PHE-PHEPHE ($M_2$), Cu-PHE-PHEOME ($M_3$) and Cu-PHE-PHEME ($M_4$) with the help of the donor and acceptor CSP lenses. \autoref{fig:substituents-lens} shows the structure of the individual molecules. In all molecules PHE ($L_1$) appears on the left and Cu in the center. The third subgroup (right) varies from top to bottom as PHE ($L_1$), PHEPHE ($L_2$), PHEOME ($L_3$) and PHEME ($L_4$) depending on the substituent atoms added to $L_1$. The aim is to study the behavior of $L_1$ with different substituents.
We extract the donor and acceptor CSPs using donor and acceptor lenses, respectively. Light green and red control polygons represent lower donor and acceptor strength as compared to the darker ones. For $M_1$ and $M_2$, fiber surfaces corresponding to both light and dark green control polygons form an envelope around the copper atom or its neighboring atoms which implies that the donor region, whether weak or strong, lies in the vicinity of the copper atom. This observation indicates that Cu is the only donor in both molecules and, to some extent, its behavior has impacted the neighborhood region. By default $L_1$ and $L_2$ are expected to be acceptors. The fiber surfaces corresponding to the red control polygons (both light and dark) are spread across $L_1$ and $L_2$ in both molecules. This implies that converting $L_1$ to $L_2$ by adding substituents does not result in a change in the donor behavior. In $M_3$ and $M_4$, we observe that a significant portion of the red fiber surfaces lie within $L_1$ and hence the sole acceptor subgroup is $L_1$. In both molecules, Cu is the major donor. One interesting observation is that after adding substituents to $L_1$, $L_3$ and $L_4$ also exhibit weak donor fiber surfaces (green) and this behavior is no longer limited to the close neighbors of Cu. This implies that substituents have impacted the complete subgroup and it is behaving as a donor, although the donor strength is much lesser than Cu. Unlike CSP peel operator, the CSP lens operator works without any prior subgroup information because the operator is applied directly in the range space. The molecular structures are shown only for context. Resulting output CSPs are linked to domain by fiber surfaces for further analysis. In this case study, each donor/acceptor CSP is linked to donor/acceptor fiber surface.

\section{Conclusions}
In this paper, we presented two CSP based operators for bivariate data analysis. The two operators provide a new direction to the analysis of bivariate fields. CSP lens operator is a simple and quick method to directly query the range space. It also simplifies and drives the control polygon selection process. The CSP peel operator helps extract layers of the CSP based on a user-specified domain segmentation. The donor strength is quantified, and depends on both NTO fields. The relationship between the fields directs the selection of control polygons rather than an interactive manual selection. We present detailed case studies that demonstrate the effectiveness of the two operators and the visual analysis pipeline. In future work, we plan to extend the analysis approach to other applications and generalize the method to multifield data.


%



\ifCLASSOPTIONcompsoc
  \section*{Acknowledgments}
\else
  \section*{Acknowledgment}
\fi

This work is partially supported by an Indo-Swedish joint network project: DST/INT/SWD/VR/P-02/2019 and VR grant 2018-07085, MoE Govt. of India, a Swarnajayanti Fellowship from DST India (DST/SJF/ETA-02/2015-16), a Mindtree Chair research grant, the SeRC (Swedish e-Science Research Center), the Swedish Research Council~(VR) grant 2019-05487. 
SST is associated to the Wallenberg AI, Autonomous Systems and Software Program (WASP).
The computations were enabled by resources provided by the Swedish National Infrastructure for Computing (SNIC) at NSC partially funded by the VR grant agreement no. 2018-05973.

\ifCLASSOPTIONcaptionsoff
  \newpage
\fi



%


\bibliographystyle{IEEEtran}
\bibliography{references}

\begin{thebibliography}{10}
\providecommand{\url}[1]{#1}
\csname url@samestyle\endcsname
\providecommand{\newblock}{\relax}
\providecommand{\bibinfo}[2]{#2}
\providecommand{\BIBentrySTDinterwordspacing}{\spaceskip=0pt\relax}
\providecommand{\BIBentryALTinterwordstretchfactor}{4}
\providecommand{\BIBentryALTinterwordspacing}{\spaceskip=\fontdimen2\font plus
\BIBentryALTinterwordstretchfactor\fontdimen3\font minus
  \fontdimen4\font\relax}
\providecommand{\BIBforeignlanguage}[2]{{%
\expandafter\ifx\csname l@#1\endcsname\relax
\typeout{** WARNING: IEEEtran.bst: No hyphenation pattern has been}%
\typeout{** loaded for the language `#1'. Using the pattern for}%
\typeout{** the default language instead.}%
\else
\language=\csname l@#1\endcsname
\fi
#2}}
\providecommand{\BIBdecl}{\relax}
\BIBdecl

\bibitem{Kim2019}
H.~W. Kim, K.~Kim, S.~W. Park, and Y.~M. Rhee, ``Quantum chemistry for studying
  electronic spectroscopy and dynamics of complex molecular systems,'' in
  \emph{Molecular Spectroscopy: A Quantum Chemistry Approach}.\hskip 1em plus
  0.5em minus 0.4em\relax Wiley Online Library, 2019, vol.~1.

\bibitem{Mulliken}
R.~S. {Mulliken}, ``{Electronic structures of polyatomic molecules and valence.
  II. General considerations},'' \emph{Physical Review}, vol.~41, no.~1, pp.
  49--71, Jul. 1932.

\bibitem{Martin2003NTO}
R.~Martin, ``Natural transition orbitals,'' \emph{J. Chem. Phys.}, vol. 118,
  pp. 4775--4777, 03 2003.

\bibitem{Bachthaler2008CSP}
S.~Bachthaler and D.~Weiskopf, ``Continuous scatterplots,'' \emph{IEEE
  Transactions on Visualization and Computer Graphics}, vol.~14, no.~6, pp.
  1428--1435, 2008.

\bibitem{carr2015fiber}
H.~Carr, Z.~Geng, J.~Tierny, A.~Chattopadhyay, and A.~Knoll, ``Fiber surfaces:
  Generalizing isosurfaces to bivariate data,'' \emph{Computer Graphics Forum},
  vol.~34, no.~3, pp. 241--250, 2015.

\bibitem{Humphrey1996}
W.~Humphrey, A.~Dalke, and K.~Schulten, ``{VMD -- Visual molecular dynamics},''
  \emph{Journal of Molecular Graphics}, vol.~14, pp. 33--38, 1996.

\bibitem{Stone2011}
J.~E. Stone, D.~J. Hardy, J.~Saam, K.~L. Vandivort, and K.~Schulten,
  ``{GPU-accelerated computation and interactive display of molecular
  orbitals},'' in \emph{GPU Computing Gems Emerald Edition}, ser. Applications
  of GPU Computing Series.\hskip 1em plus 0.5em minus 0.4em\relax Elsevier
  Inc., 2011, pp. 5--18.

\bibitem{Haranczyk2008}
M.~Haranczyk and M.~Gutowski, ``{Visualization of molecular orbitals and the
  related electron densities},'' \emph{Journal of Chemical Theory and
  Computation}, vol.~4, no.~5, pp. 689--693, 2008.

\bibitem{Garcia2010}
G.~Garcia, C.~Adamo, and I.~Ciofini, ``{Evaluating push--pull dye efficiency
  using TD-DFT and charge transfer indices},'' \emph{Phys. Chem. Chem. Phys.},
  vol.~15, pp. 20\,210--20\,219, 2010.

\bibitem{Bahers2011}
T.~L. Bahers, C.~Adamo, and I.~Ciofini, ``{A qualitative index of spatial
  extent in charge-transfer excitations},'' \emph{J Chem. Theory Comput.},
  vol.~7, pp. 2498--2506, 2011.

\bibitem{Guido2013}
C.~A. Guido, P.~Cortona, B.~Mennucci, and C.~Adamo, ``{On the metric of charge
  transfer molecular excitations: A simple chemical descriptor},''
  \emph{Journal of Computational Chemistry}, vol.~9, no.~7, pp. 3118--3126,
  2013.

\bibitem{Huet2020}
L.~Huet, A.~Perfetto, F.~Muniz-Miranda, M.~Campetella, C.~Adamo, and
  I.~Ciofini, ``{General density-based index to analyze charge transfer
  phenomena: From models to butterfly molecules},'' \emph{J. Chem. Theory
  Comput.}, vol.~16, pp. 4543--4553, 2020.

\bibitem{masood2021visual}
T.~B. Masood, S.~S. Thygesen, M.~Linares, A.~I. Abrikosov, V.~Natarajan, and
  I.~Hotz, ``{Visual analysis of electronic densities and transitions in
  molecules},'' \emph{Computer Graphics Forum}, vol.~40, no.~3, 2021.

\bibitem{Thygesen2022}
S.~S. Thygesen, T.~B. Masood, M.~Linares, V.~Natarajan, and I.~Hotz, ``{Level
  of detail exploration of electronic transition ensembles using hierarchical
  clustering},'' \emph{Computer Graphics Forum}, vol.~41, no.~3, pp. 333--344,
  2022.

\bibitem{sarikaya2017scatterplots}
A.~Sarikaya and M.~Gleicher, ``{Scatterplots: Tasks, data, and designs},''
  \emph{IEEE Transactions on Visualization and Computer Graphics}, vol.~24,
  no.~1, pp. 402--412, 2017.

\bibitem{Lehmann2010}
D.~J. Lehmann and H.~Theisel, ``{Discontinuities in continuous scatter
  plots},'' \emph{IEEE Transactions on Visualization and Computer Graphics},
  vol.~16, no.~6, pp. 1291--1301, 2010.

\bibitem{nagaraj2010relation}
S.~Nagaraj and V.~Natarajan, ``{Relation-aware isosurface extraction in
  multifield data},'' \emph{IEEE Transactions on Visualization and Computer
  Graphics}, vol.~17, no.~2, pp. 182--191, 2010.

\bibitem{hamish2006histograms}
H.~Carr, B.~Duffy, and B.~Denby, ``{On histograms and isosurface statistics},''
  \emph{IEEE Transactions on Visualization and Computer Graphics}, vol.~12,
  no.~5, pp. 1259--1266, 2006.

\bibitem{klacansky2016fast}
P.~Klacansky, J.~Tierny, H.~Carr, and Z.~Geng, ``{Fast and exact fiber surfaces
  for tetrahedral meshes},'' \emph{IEEE Transactions on Visualization and
  Computer Graphics}, vol.~23, no.~7, pp. 1782--1795, 2016.

\bibitem{tierny2016jacobi}
J.~Tierny and H.~Carr, ``{Jacobi fiber surfaces for bivariate {R}eeb space
  computation},'' \emph{IEEE Transactions on Visualization and Computer
  Graphics}, vol.~23, no.~1, pp. 960--969, 2016.

\bibitem{Blecha2019Nuclear}
C.~Blecha, F.~Raith, G.~Scheuermann, T.~Nagel, O.~Kolditz, and J.~Maßmann,
  ``{Analysis of coupled thermo-hydro-mechanical simulations of a generic
  nuclear waste repository in clay rock using fiber surfaces},'' in \emph{2019
  IEEE Pacific Visualization Symposium (PacificVis)}, 2019, pp. 189--201.

\bibitem{Raith2019Tensor}
F.~Raith, C.~Blecha, T.~Nagel, F.~Parisio, O.~Kolditz, F.~Günther, M.~Stommel,
  and G.~Scheuermann, ``{Tensor field visualization using fiber surfaces of
  invariant space},'' \emph{IEEE Transactions on Visualization and Computer
  Graphics}, vol.~25, no.~1, pp. 1122--1131, 2019.

\bibitem{edelsbrunner2008reeb}
H.~Edelsbrunner, J.~Harer, and A.~K. Patel, ``{Reeb spaces of piecewise linear
  mappings},'' in \emph{Proceedings of the twenty-fourth annual symposium on
  Computational geometry}, 2008, pp. 242--250.

\bibitem{sharma2021segmentation}
M.~Sharma, T.~B. Masood, S.~S. Thygesen, M.~Linares, I.~Hotz, and V.~Natarajan,
  ``{Segmentation driven peeling for visual analysis of electronic
  transitions},'' in \emph{Proc. {IEEE} Visualization Conference, {IEEE} {VIS}
  2021 - Short Papers}.\hskip 1em plus 0.5em minus 0.4em\relax IEEE, 2021, pp.
  96--100.

\bibitem{Tierny2018ttk}
J.~Tierny, G.~Favelier, J.~A. Levine, C.~Gueunet, and M.~Michaux, ``{The
  topology toolKit},'' \emph{IEEE Transactions on Visualization and Computer
  Graphics}, vol.~24, no.~1, pp. 832--842, 2018.

\bibitem{scheidegger2008revisiting}
C.~E. Scheidegger, J.~M. Schreiner, B.~Duffy, H.~Carr, and C.~T. Silva,
  ``{Revisiting histograms and isosurface statistics},'' \emph{IEEE
  Transactions on Visualization and Computer Graphics}, vol.~14, no.~6, pp.
  1659--1666, 2008.

\bibitem{Aurenhammer1987powerdiag}
F.~Aurenhammer, ``{Power diagrams: Properties, algorithms and applications},''
  \emph{SIAM J. Comput.}, vol.~16, no.~1, p. 78–96, 1987.

\bibitem{Ayachit2015paraview}
U.~Ayachit, \emph{{The ParaView Guide, A Parallel Visualization Application}},
  2015.

\bibitem{morgan2000geometric}
F.~Morgan, \emph{{Geometric Measure Theory: A Beginner’s Guide}}, 2000.

\bibitem{Frisch2016Gaussian}
M.~J. Frisch, G.~W. Trucks, H.~B. Schlegel, G.~E. Scuseria, and {M. A. Robb et
  al.}, ``Gaussian~16,'' 2016, {Gaussian Inc. Wallingford CT}.

\end{thebibliography}

%

\vskip -2.2\baselineskip plus -1fil
\begin{IEEEbiography}[{\includegraphics[width=1in,height=1.25in,clip,keepaspectratio]{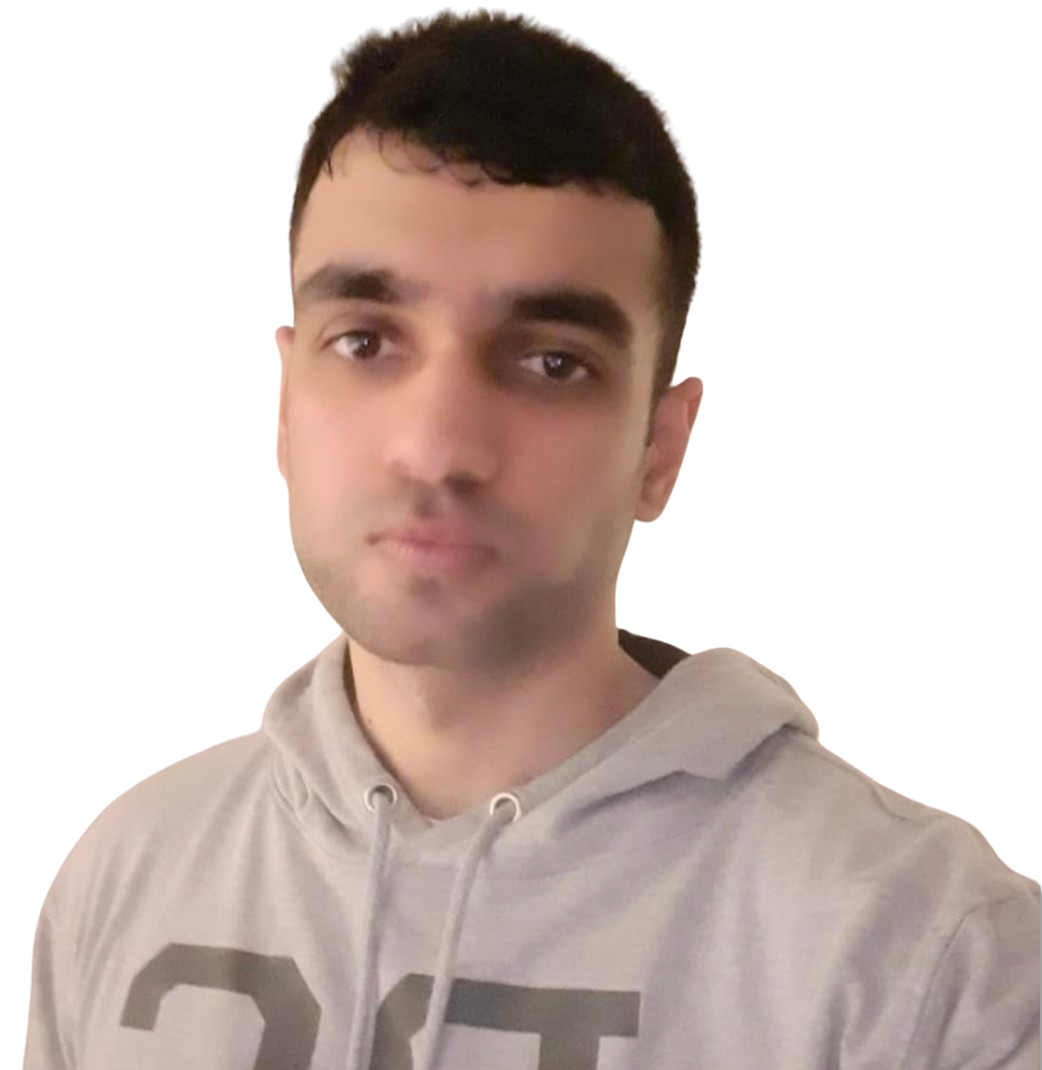}}]{Mohit Sharma}
is a Ph.D. student in the Computer Science and Automation Department at Indian Institute of Science, Bangalore. He received his M.Tech degree in Computer Science and Engineering from IIIT, Hyderabad. His research interests include scientific visualization, computational topology and its applications. Currently, he is working on multi-field data visualization.
\end{IEEEbiography}

\vskip -2.2\baselineskip plus -1fil
\begin{IEEEbiography}[{\includegraphics[width=1in,height=1.25in,clip,keepaspectratio]{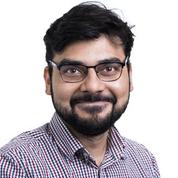}}]{Talha Bin Masood} is a Postdoctoral Fellow at Link\"{o}ping University in Sweden. 
He received his Ph.D. in Computer Science from the Indian Institute  of  Science, Bangalore.  His research interests  include scientific  visualization, computational  geometry, computational topology, and their applications to various scientific domains.\end{IEEEbiography}


\vskip -2.2\baselineskip plus -1fil
\begin{IEEEbiography}[{\includegraphics[width=1in,height=1.25in,clip,keepaspectratio]{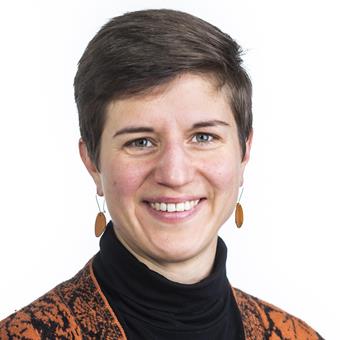}}]{Signe S. Thygesen}
is a Ph.D. student in Scientific Visualization at Linköping University in Sweden. She graduated with a Master of Science degree in Engineering Physics from Lund University. Her research interests include scientific visualization and topological data analysis with applications to scientific domains. In recent work, she has developed methods for exploration and visual representation of molecular electronic transitions.
\end{IEEEbiography}

\vskip -2.2\baselineskip plus -1fil
\begin{IEEEbiography}[{\includegraphics[width=1in,height=1.25in,clip,keepaspectratio]{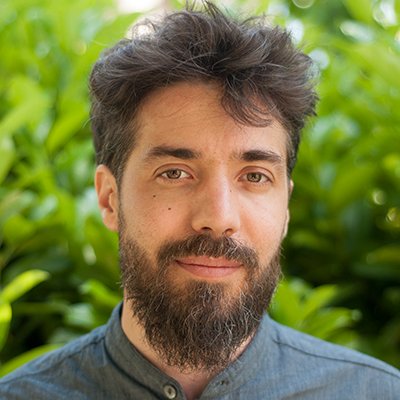}}]{Mathieu Linares}
obtained
his PhD in Theoretical Chemistry at the Paul Cézanne University (Marseilles, France) in 2005,
where he developed a valence bond method. In 2006,
he joined the Laboratory for Chemistry of Novel
Materials in Mons (Belgium) for a post-doctoral
stay where he worked on the modeling of self-assembly
in solution and at surfaces. He then
worked as a postdoctoral researcher in the Computational
Physics group at Link\"{o}ping University (2008-2009) and in the
group of Theoretical Chemistry at the Royal Institute of Technology, Stockholm, Sweden (2010). He currently holds
an Associate Professor position shared between the laboratory of organic
electronics and the group of scientific visualization at Link\"{o}ping
University, Sweden. His main research interests are chirality,
self-assembly in solution and on surface, opto-electronic properties of organic
materials, and molecular visualization.
\end{IEEEbiography}

\vskip -2.2\baselineskip plus -1fil
\begin{IEEEbiography}[{\includegraphics[width=1in,height=1.25in,clip,keepaspectratio]{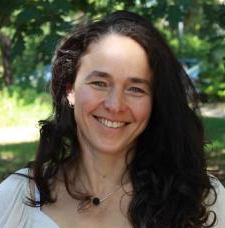}}]{Ingrid Hotz}
is currently a Professor in Scientific Visualization at the Link\"{o}ping University in Sweden. She received her Ph.D. degree from the Computer Science Department at the University of Kaiserslautern, Germany. Her research interests lie in data analysis  and  scientific  visualization,  ranging  from  basic  research questions  to  effective  solutions  to  visualization  problems  in  applications.  
\end{IEEEbiography}

\vskip -2.2\baselineskip plus -1fil
\begin{IEEEbiography}[{\includegraphics[width=1in,height=1.25in,clip,keepaspectratio]{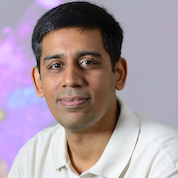}}]{Vijay Natarajan} is the Mindtree Chair Professor in the Department of Computer Science and Automation at Indian Institute of Science, Bangalore. He received the Ph.D. degree in computer science from Duke University. His research interests include scientific visualization, computational topology, and computational geometry. In current work, he is developing topological methods for time-varying and multi-field data visualization, and studying applications in biology, material science, and climate science.
\end{IEEEbiography}




\end{document}